\documentclass[a4paper,USenglish]{lipics-v2016}

\usepackage{amsmath}
\usepackage{microtype}

\title{P2P-PL: A Pattern Language to Design Efficient and Robust Peer-to-Peer Systems}
\titlerunning{P2P-PL}

\author{Michele Amoretti}
\author{Francesco Zanichelli}
\affil{Department of Information Engineering, Universit\`a degli Studi di Parma, Italy\\
  \texttt{\{michele.amoretti,francesco.zanichelli\}@unipr.it}}
\authorrunning{M. Amoretti and F. Zanichelli} 

\Copyright{Michele Amoretti and Francesco Zanichelli}


\begin{document}

\maketitle

\begin{abstract}
To design peer-to-peer (P2P) software systems is a challenging task, because of their highly decentralized nature, which may cause unexpected emergent global behaviors. The last fifteen years have seen many P2P applications to come out and win favor with millions of users. From success histories of applications like BitTorrent, Skype, MyP2P we have learnt a number of useful design patterns. Thus, in this article we present a P2P pattern language (shortly, P2P-PL) which encompasses all the aspects that a fully effective and efficient P2P software system should provide, namely consistency of stored data, redundancy, load balancing, coping with asymmetric bandwidth, decentralized security. The patterns of the proposed P2P-PL are described in detail, and a composition strategy for designing robust, effective and efficient P2P software systems is proposed.
\end{abstract}

\section{Introduction}
\label{intro}
Over the last fifteen years, peer-to-peer (P2P) computing has become explosively popular, particularly for massive file sharing \cite{Durhander2011,Gao2012,Belmonte2013} and multimedia Internet streaming \cite{Cameron2014,Jia2014,Kao2016}, whereas very high scalability has also been sought more recently in additional application domains, namely online gaming and location-aware services.

As the P2P paradigm is gaining a forefront position in distributed computing techniques \cite{Rius2013}, it is important to recap good practices in P2P software  system design and express them in terms of pattern language. Before reviewing previous P2P pattern languages and introducing our proposal for a P2P-PL, we deem necessary to illustrate the essential properties of a P2P system.

A P2P network is a complex software system whose elements (peer nodes, or simply peers) cooperate by partitioning tasks and workloads, in order to ensure correctness, efficiency and robustness with invariance of scale. In the most radical view of P2P computing, peers participate in the application with equal privileges. Peers are said to form an overlay network of nodes, on top of the underlying computer (typically IP) network \cite{Schoder04}. 

Such a (well recognized) definition implies that peers share their \textit{resources}: CPU, storage, bandwidth, cache, files, applications, services. Sharing implies that resources are discoverable. Some peers are greedy but share a minimal quantity of resources, which is why they are denoted as \textit{free riders}. Usually, all peers implicitly provide \textit{basic shared resources} (\textit{e.g.}, storage and bandwidth). We denote as \textit{consumable resources} those that cannot be obtained by replication and can only be exploited upon contracting with their hosts. Usually, a limited number of peers can concurrently access the consumable resources of a peer. Services are special resources that can be grouped in two categories: \textit{distributed services}, whose execution involves several peers, each one contributing with its basic shared resources, and \textit{local services}, which are isolated functional units allowing to access the local non-basic resources of a peer. Local services may also be denoted as \textit{resource provision services}.

Resource sharing mechanisms in P2P software systems (P2P systems, henceforth) can be either fully decentralized or supported by central controllers. Although P2P purists only accept total decentralization, it is a matter of fact that several renowned P2P networks (\textit{e.g.}, Napster, eMule, BitTorrent) are based on a centralized resource index.

P2P characterizing properties such as scalability and load balancing should be a consequence of proper decisions in the design of the connectivity layer, which may bring to a more or less structured overlay topology. The choice of the topology and routing strategy are often strongly interwoven. This interdependency complicates the design of efficient P2P systems. Other key problems, known since the beginning of the P2P era \cite{WilcoxOHearn2002}, are bypassing firewalls and NAT, attack resistance, mutual distrust and improving the motivation to cooperate.  

To solve these and other issues, a suitable pattern language would greatly benefit the designer, as stated, among others, by Buschmann \textit{et al.} \cite{Buschmann2007}. However, the definition of a P2P pattern language has received limited attention so far in the literature.

The EuroPlop 2002 focus group on P2P patterns \cite{EP2002} discussed on the most important technical characteristics of P2P systems and on the issues related to designing and building them. Although no patterns were defined, it was remarked that characterizing properties of P2P should include dynamic service relationship management, distributed state, multi-hop routing with recovery, adaptive reconfiguration, coordination and load-balancing, as well as fault-tolerance and security.

Alesky et al. \cite{Aleksy06} analyzed seven well-known P2P systems to determine which design patterns are adopted. Results (summarized in Table \ref{tab:mostUsedPatterns}) are not surprising: creational, structural \cite{GoF}, and concurrency/networking \cite{POSA2} patterns are used.

\begin{center}
\begin{table}[!h]
    \caption {Most used basic design patterns in P2P architectures \cite{Aleksy06}} 
    \label{tab:mostUsedPatterns} 
    \begin{tabular}{ | p{3cm} | p{0.8cm} | p{6cm} | }
    \hline
   Pattern & \# & P2P Systems \\ \hline \hline
   \textbf{Observer} & 5 & Limewire, Gridella, Frost, FSS, ProActive \\ \hline
    \textbf{Active Object} & 4 & Limewire, Gridella, Phex, ProActive \\ \hline
    \textbf{Singleton} & 4 & Limewire, Gridella, Frost, ProActive \\ \hline
    \textbf{Caching} & 4 & Frost, FSS, JXTA, Phex \\ \hline
    \textbf{Acceptor-Connector} & 3 & Limewire, Gridella, Phex \\ \hline
    \textbf{Non Blocking I/O} & 3 & Limewire, Gridella, Phex \\ \hline
    \textbf{Asynchronous Completion Token} & 3 & Limewire, Gridella, Phex \\ \hline
    \textbf{Location} & 3 & FSS, JXTA, ProActive \\ \hline
    \textbf{Monitored Object} & 3 & Limewire, Gridella, Phex \\ \hline
    \textbf{Proxy} & 3 & Limewire, FSS, ProActive \\ \hline
    \textbf{Wrapper Facade} & 3 & Limewire, JXTA, Phex \\ \hline
    \textbf{Facade} & 3 & Limewire, Gridella, ProActive \\ \hline
   \end{tabular}
\end{table}
\end{center}

\begin{figure}[!h]
\centering
\includegraphics[width=12cm]{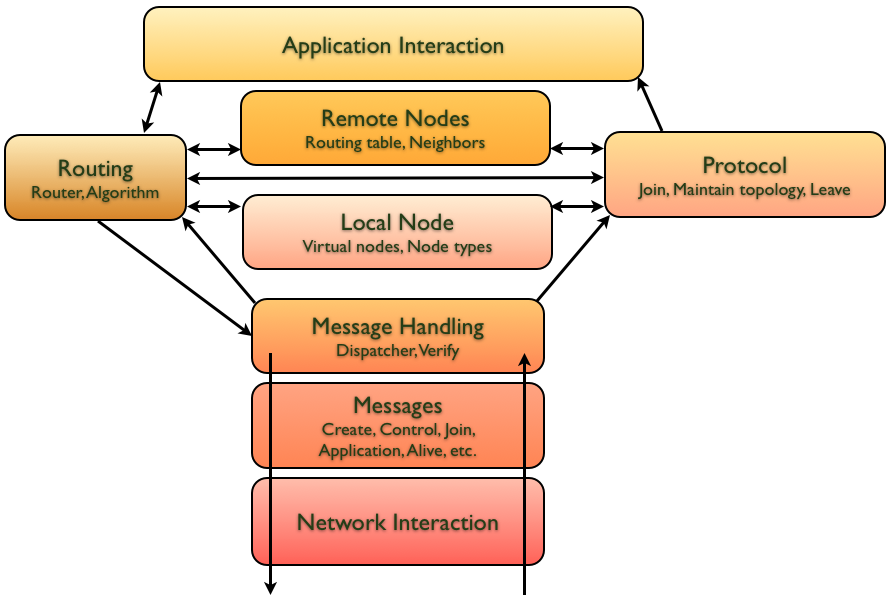}
\caption{The pattern language presented by Grolimund and Muller \cite{Grolimund06}.}
\label{fig:grolimund}
\end{figure}

The pattern language proposed by Grolimund and Muller \cite{Grolimund06} consists of adaptations of existing patterns (\textit{e.g.}, \textbf{Facade} and \textbf{Observer} \cite{GoF}) as well as several new proto-patterns belonging to the core building blocks of overlay networks. Such proto-patterns are related to application interaction, messages and message handling, routing protocol, as well as network interaction. The pattern language introduced by Grolimund and Muller is schematically depicted in Figure \ref{fig:grolimund}, where a bottom-up perspective can be noted.

In a previous work, we presented an architectural design pattern, called \textbf{Peer} \cite{Amoretti2005a}, aimed at capturing the core properties of a generic P2P system. 
Since then, the set of features that may characterize such a system has grown exponentially, well beyond what a single pattern can express, thus making a pattern language necessary.
Based on our experience on P2P systems, from modeling \cite{Amoretti2012a} and performance characterization \cite{Amoretti2006} to middleware development (formerly with SP2A \cite{Amoretti2005b}, and now with NAM4J \cite{Amoretti2012b}), we present a \textit{P2P Pattern Language} (shortly, \textit{P2P-PL}) encompassing the main aspects that a state-of-the-art P2P architecture should provide, including consistency of stored data, redundancy, load balancing, coping with asymmetric bandwidth, decentralized security.

In writing the P2P-PL, we have followed the approach suggested by the \textbf{Pattern Language for Pattern Writing} \cite{Meszaros1997}, and we have put much effort in order to satisfy the pattern quality criteria described by Wurhofer \textit{et al.} \cite{Wurhofer2010}. The first criterion, called \textit{findability}, states that a pattern must be found easily and quickly within a pattern collection or pattern language. P2P-PL patterns are organized into a layered structure, which makes it easy to distinguish between core patterns and accessory ones. The second criterion, \textit{understandability}, deals with the fact that the pattern must be easily understood by its users. To this purpose, P2P-PL patterns
\begin{itemize}
\item have meaningful names; 
\item contain all relevant descriptions of forces, problems, solutions and examples;
\item use a language that is easy to understand;
\item are centered around a problem.
\end{itemize}
With respect to \textit{helpfulness}, P2P-PL patterns provide sufficient information about how to implement them.
The fourth criterion, \textit{empirical verification}, describes the fact that a pattern must be approved by empirical data. All P2P-PL patterns have been implemented (with several variations) into real P2P systems, most of them being highly popular because of their effectiveness and efficiency.  
For the same reason, P2P-PL patterns meet also the fifth criterion, namely \textit{overall acceptability}.

The paper is organized as follows. In Section \ref{pattLangSummary}, the main features of P2P-PL are summarized, considering the problem and related solution characterizing each pattern, as well as a layered architecture, a dependency diagram and a ``story''. In Section \ref{patterns}, all P2P-PL patterns are illustrated in detail, one by one. In Section \ref{implementation}, an effective procedure for using P2P-PL to implement P2P systems and the BitTorrent example are presented. Finally, some concluding remarks are proposed in Section \ref{conclusions}.

\section{Pattern Language Summary}
\label{pattLangSummary}

P2P-PL is based on a set of \textit{architectural patterns} that cover a wide range of P2P-specific features. 
Our top-down analysis has evidenced recurring structural and behavioral elements existing in almost all successful P2P applications, across different application domains (such as file sharing, Internet streaming, collaborative work and ambient intelligence).  
P2P-PL complements the pattern language proposed by Grolimund and Muller \cite{Grolimund06} as it emphasizes most of the advanced global properties and behavior of state-of-the-art P2P systems, and relates them especially to the overlay organization.
Usually, the design of a peer's internal organization is also based on patterns for general distributed systems, such as the \textbf{Remoting Patterns} illustrated by Voelter \textit{et al.} \cite{Voelter04}.

Table \ref{tab:overview} lists the design problems addressed by P2P-PL, and summarizes the corresponding solutions provided by the pattern language itself.

\begin{center}
\footnotesize
\begin{table}
    \caption {Overview of the patterns in P2P-PL} 
    \label{tab:overview} 
     \footnotesize
    \begin{tabular}{ | p{4cm} | p{5cm} | p{4cm} | }
    \hline
    Problem & Solution & Pattern Name \\ \hline \hline
    How do the components of the peer-to-peer system organize themselves as a whole system and interact? & Organize the peers according to an \textbf{Overlay Scheme}, \textit{i.e.}, connect them by virtual links, corresponding to paths in the underlying network. & \textbf{Overlay Scheme} \\ \hline
    How to make joining peers join the overlay network? & Allow joining peers to obtain a list of peers which should be online, to let them virtually link to at least one peer that is already a member of the overlay network. & \textbf{Bootstrapping} \\ \hline
    How to model the topology of the overlay network? & Describe the topology in terms of distribution of the node degree, clustering coefficient, average connected distance and diameter. & \textbf{Topology} \\ \hline
    How are messages propagated in the overlay network? & Define a component that consumes messages from a message channel. Consumed messages are forwarded each one to a different message channel, depending on a set of conditions. & \textbf{Message Routing} \\ \hline
    How to name, put, replicate and get resources within the overlay network? & Let each resource have a unique identifier, and provide peers with PUT, STORE, GET, LOOKUP mechanisms. & \textbf{Distributed State} \\ \hline
    How to preserve the consistency of resource descriptors within the overlay network? & Choose the most convenient place to store resource descriptors, which should have a limited lifetime and should be periodically refreshed. Also, allow requesters to keep track of resource owners. & \textbf{Information Consistency} \\ \hline
    How to guarantee the integrity, authenticity and availability of data stored in the P2P system? & Introduce a voting mechanism, sign data, replicate data either actively or passively, adopt secure routing policies in order to prevent malicious nodes polluting forwarded data. & \textbf{Data Protection} \\ \hline
    How to avoid issues due to open or loosely controlled membership? & Create groups of peers with common interests and capabilities. & \textbf{Group Membership} \\ \hline
    How to improve the performance of data transfer among peers? & Simultaneously download data from multiple sources. & \textbf{Multisource Data Transfer} \\ \hline
    How to guarantee a reasonable degree of data upload and download mutuality among peers? & Adopt a peer selection strategy that penalizes free riders, \textit{i.e.}, peers that download but never upload data. & \textbf{Choke/Unchoke} \\ \hline
    How to reward the peers that contribute to the functioning of the network, while penalizing free riders?  & Adopt a peer reputation scheme. & \textbf{Reputation} \\ \hline
    \end{tabular}
\end{table}
\end{center}

\begin{figure}[!h]
\centering
\includegraphics[width=12cm]{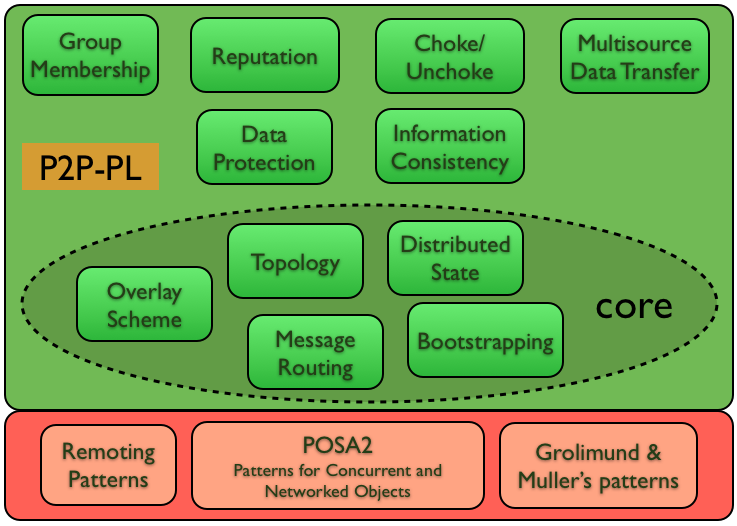}
\caption{Layered vision of the P2P-PL presented in this paper.}
\label{fig:layers}
\end{figure}

Figure \ref{fig:layers} illustrates how previously listed patterns are placed with respect to the other, more basic design patterns suggested by Alesky et al. \cite{Aleksy06} and Grolimund and Muller \cite{Grolimund06}.

\begin{figure}[!h]
\centering
\includegraphics[width=12cm]{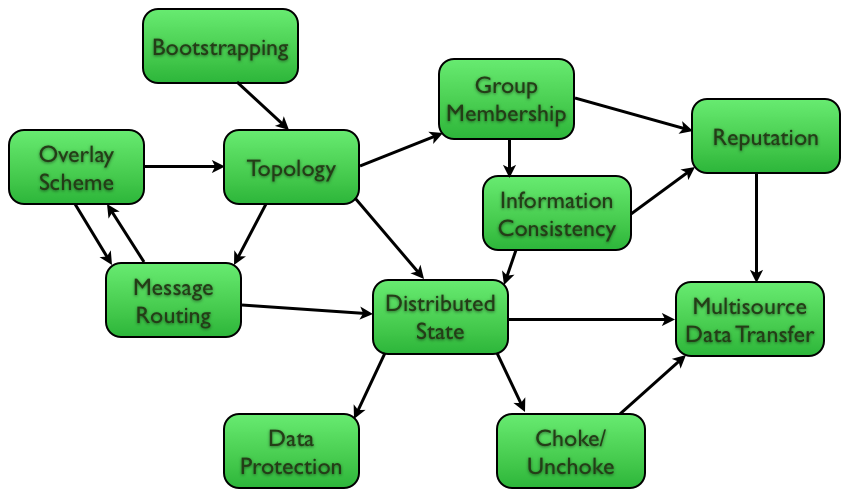}
\caption{Dependency diagram of the P2P-PL presented in this paper.}
\label{fig:depend}
\end{figure}

Before illustrating each pattern in detail, we show how P2P-PL subsets can be used to implement different P2P systems. Firstly, we need to define how P2P-PL patterns depend on each other, by means of a dependency graph (illustrated in Figure \ref{fig:depend}). It can be observed that \textbf{Overlay Scheme} and \textbf{Message Routing} have mutual dependences, as it is not always possible to design separately and independently the organization of the peers and the way they share information.
As previously stated, \textbf{Overlay Scheme}, \textbf{Topology}, \textbf{Message Routing}, \textbf{Bootstrapping} and \textbf{Distributed State} patterns are mandatory to build any P2P architecture. Therefore, they are the core of P2P-PL. Whether other P2P-PL patterns are necessary or not, it depends on the specific P2P application.

The P2P-PL allows to tell the following general ``story'':

\begin{quote}
\textit{
Every peer-to-peer system is characterized by an \textbf{Overlay Scheme}, which defines the relationship between the peers' reciprocal knowledge, modeled as a dynamic \textbf{Topology}, and the \textbf{Message Routing} strategy they adopt. To join the overlay network, each peer has to perform a \textbf{Bootstrapping} process. Resource identification mechanisms allow to be in touch with the \textbf{Distributed State} of the system, whose reliability must be supported by adequate policies for guaranteeing \textbf{Information Consistency} and \textbf{Data Protection}. For security reasons or application scoping, peers may adopt \textbf{Group Membership} strategies. Their \textbf{Reputation} may be a good means for deciding whether to allow or prevent them to access resources and services. For example, in a \textbf{Multisource Data Transfer} process, peers with bad \textbf{Reputation} may be penalized by a \textbf{Choke/Unchoke} mechanism that decides download turns. 
}
\end{quote}

The latter story can be further specialized, with reference to different P2P applications. Table \ref{tab:P2Papps} shows, for the main types of P2P applications, which P2P-PL patterns may apply. As core P2P-PL patterns are tacit for every application domain, they are not explicitly listed. The remaining patterns are not all mandatory. For example, the BitTorrent content sharing network lacks \textbf{Information Consistency}, thus one may find descriptors of files which are no more available in the system.

\begin{center}
\begin{table*}
\caption {Association between P2P-PL patterns and P2P applications. Core P2P-PL patterns are tacit for every application domain.} 
\label{tab:P2Papps} 
\begin{tabular}{ | p{5cm} | p{6cm} |}
\hline
\hline
\textbf{P2P application domain} & \textbf{Suggested P2P-PL patterns}\\
\hline\hline
Content Sharing & \textbf{Information Consistency}, \textbf{Group Membership}, \textbf{Multisource Data Transfer}, \textbf{Choke/Unchoke}, \textbf{Reputation}  \\
\hline
Distributed Storage & \textbf{Information Consistency}, \textbf{Data Protection}, \textbf{Reputation} \\
\hline
Parallel \& Distributed Computing & \textbf{Multisource Data Transfer}, \textbf{Information Consistency} \\
\hline
Business & \textbf{Information Consistency}, \textbf{Data Protection}, \textbf{Group Membership}, \textbf{Reputation} \\
\hline
VoIP \& Multimedia Streaming & \textbf{InformationConsistency}, \textbf{Group Membership}, \textbf{Multisource Data Transfer}, \textbf{Choke/Unchoke}, \textbf{Reputation} \\
\hline
Gaming & \textbf{Information Consistency}, \textbf{Data Protection}, \textbf{Group Membership}, \textbf{Reputation} \\
\hline
Education \& Academia & \textbf{Information Consistency}, \textbf{Data Protection}, \textbf{Group Membership}, \textbf{Multisource Data Transfer}, \textbf{Reputation} \\
\hline
Ambient Intelligence & \textbf{Information Consistency}, \textbf{Data Protection}, \textbf{Group Membership} \\
\hline
\end{tabular}
\end{table*}
\end{center}

\section{Patterns} 
\label{patterns}

\subsection{\textbf{Overlay Scheme}}

\subsubsection{Context}
Early stage in the design of a peer-to-peer system, whose components (the peers) are distributed software applications that must cooperate to provide users with highly scalable global services (such as \textit{advertizing} and \textit{discovery} of peers and resources). 

\subsubsection{Problem} 
How do the components of the peer-to-peer system organize themselves as a whole system and interact?

\subsubsection{Forces}
\begin{enumerate}
\item Depending on the application(s) provided by the P2P system, the topology and the messaging strategy of the peers may be integrated or independent.
\item Peers may either all have the same role, or different roles, depending on the nature of the services provided by the P2P system.
\item The interaction scheme should take into account the (possible) different hardware capabilities of peers.
\item A well-designed P2P system should not be badly affected (in terms of effectiveness and performance) by churn, \textit{i.e.}, disconnections and (re)connections.
\end{enumerate}

\subsubsection{Solution} 
\begin{figure}[h!]
\centering
\includegraphics[width=10cm]{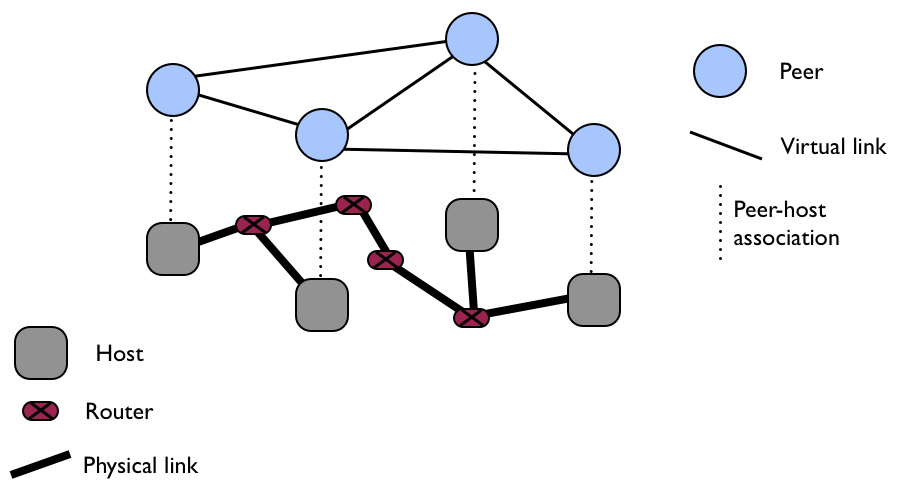}
\caption{An instance of \textbf{Overlay Scheme}. The components of the pattern are listed on the right side of the figure.}
\label{fig:overlayScheme}
\end{figure}

Organize the peers according to an \textbf{Overlay Scheme}, \textit{i.e.}, connect them by virtual links, corresponding to paths in the underlying network (as shown in Figure \ref{fig:overlayScheme}). The components of the \textbf{Overlay Scheme} pattern are the peers, the virtual links and the associations between each peer and its host. 

\subsubsection{Variations}
\begin{figure}[h!]
\centering
\includegraphics[width=6cm]{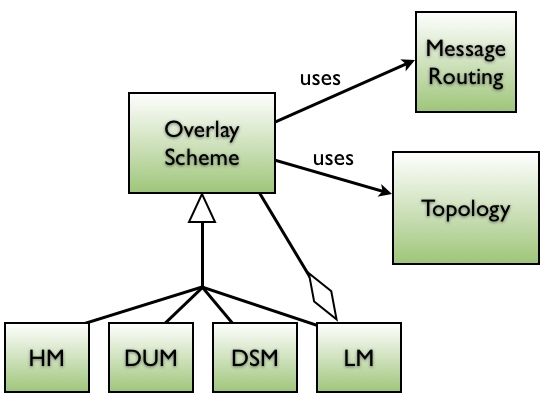}
\caption{Variations of the \textbf{Overlay Scheme} pattern.}
\label{fig:varOverlayScheme}
\end{figure}

As illustrated in Figure \ref{fig:varOverlayScheme}, one may choose among four alternative schemes: partially centralized (hybrid), decentralized unstructured, decentralized structured, or layered (hierarchical). The selection depends on how information about shared resources must be placed. In detail, information may be:
\begin{itemize}
\item placed in a central server,
\item published to other peers,
\item not published, just locally stored by owners.
\end{itemize} 

\begin{figure}[h!]
\centering
\includegraphics[width=10cm]{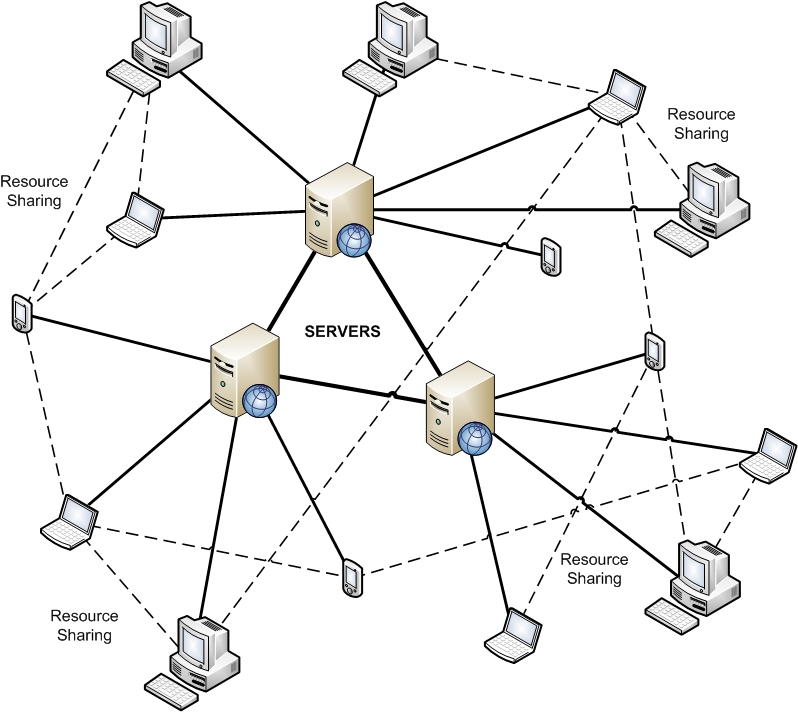}
\caption{Example of P2P system based on the Hybrid Model (HM). Peer-to-server and server-to-server reciprocal knowledge and exchange of knowledge about shared resources are represented by continuous lines. P2P resource sharing is represented by dashed lines.}
\label{fig:overlayHM}
\end{figure}

Following the first approach, \textit{hybrid} \textbf{Overlay Scheme}s (\textbf{Hybrid Model - HM}) adopt the client/server paradigm to publish and discover resources, and peer-to-peer protocols for consuming resources. An example of HM-based system is illustrated in Figure \ref{fig:overlayHM}.

\begin{figure}[h!]
\centering
\includegraphics[width=10cm]{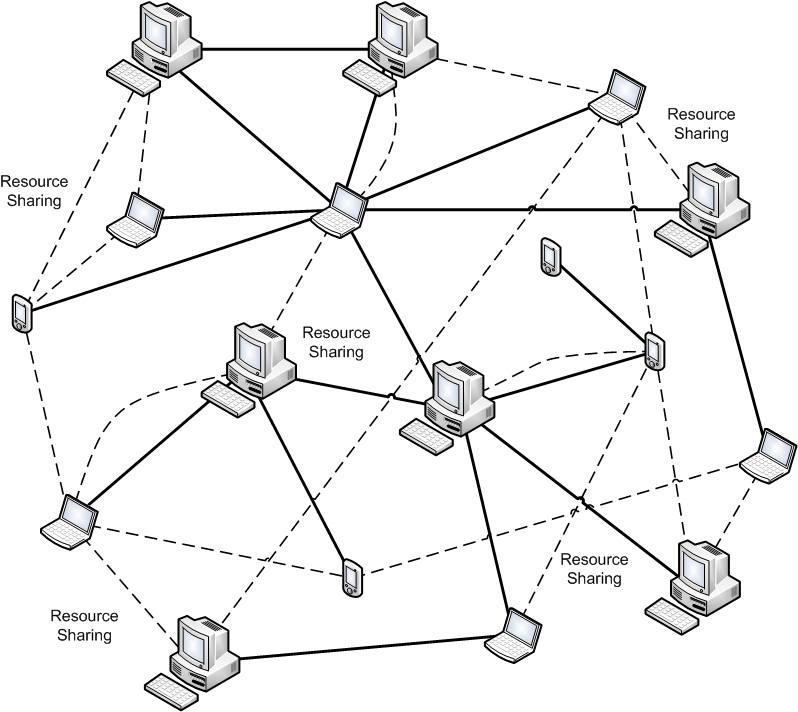}
\caption{P2P system based on the Decentralized Unstructured Model (DUM).}
\label{fig:overlayDUM}
\end{figure}

\begin{figure}[h!]
\centering
\includegraphics[width=12cm]{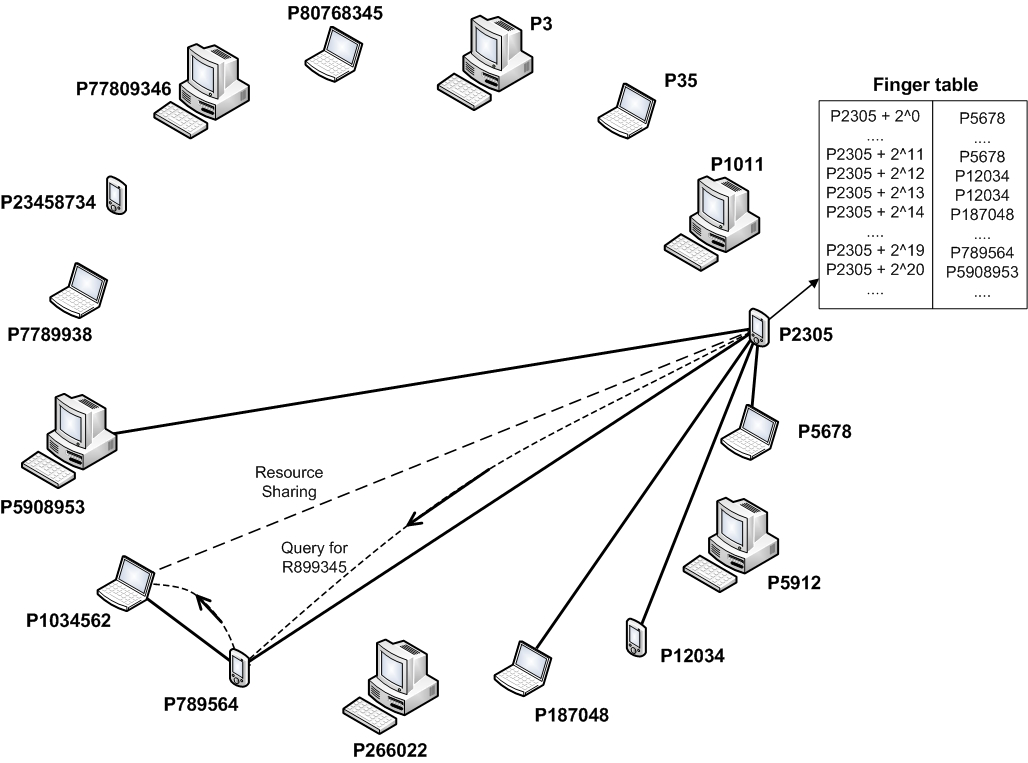}
\caption{P2P system based on the Decentralized Structured Model (DSM).}
\label{fig:overlayDSM}
\end{figure}

The other approaches define \textit{decentralized} \textbf{Overlay Scheme}s, exploiting information available locally to each node only. Resulting networks are often denoted as ``pure'' P2P systems. Depending on how much are the peers aware of the network topology, decentralized P2P systems can be further divided in two categories. We say that a decentralized P2P overlay is \textit{unstructured} (\textbf{Decentralized Unstructured Model - DUM}) if potential or actual connections among peers can be characterized as links in a random graph, whose features are not relevant to message routing strategies (Figure \ref{fig:overlayDUM}). On the contrary, a decentralized P2P overlay is denoted as \textit{structured} (\textbf{Decentralized Structured Model - DSM}) if its topology is shaped by deterministic rules and organized in a way that transferable resources (or resource advertisements) are placed at suitable locations. Each peer has a small routing table, targeting a few other (deterministically chosen) peers. Such a routing table supports publication and lookup protocols. If the publication strategy is fair enough, resources (or resource advertisements) are evenly distributed among peers, each one having the same amount of responsibility (Figure \ref{fig:overlayDSM}).

\begin{figure}[h!]
\centering
\includegraphics[width=10cm]{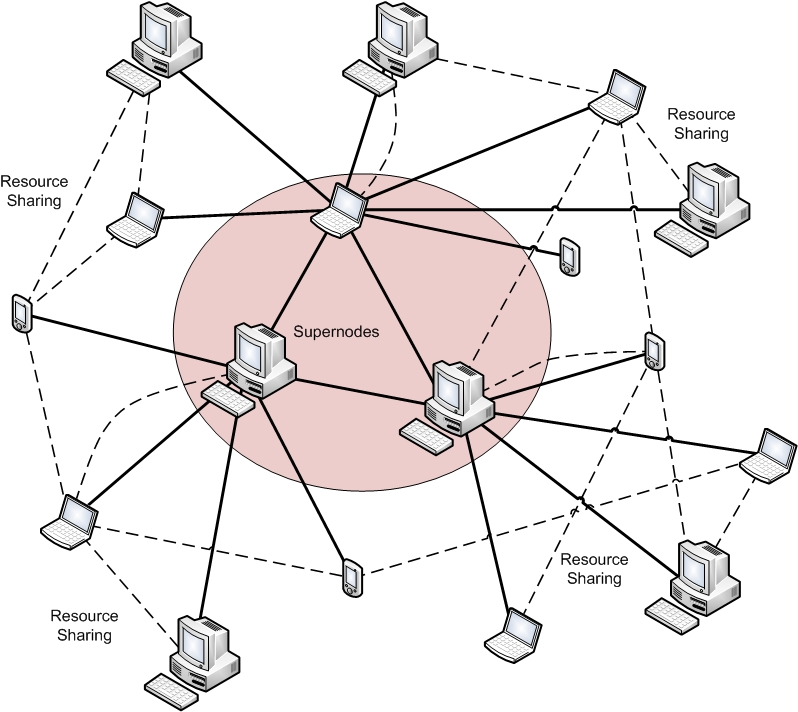}
\caption{P2P system based on a $2$-layered model.}
\label{fig:overlayLM}
\end{figure}

To obtain more robust, efficient and scalable P2P networks, \textit{layered} \textbf{Overlay Scheme}s (based on the \textbf{Layered Model - LM}) have been proposed and implemented. LM overlays (Figure \ref{fig:overlayLM}) are characterized by interacting HM, DUM, or DSM layers.

Over the last fifteen years, several different P2P architectures and protocols have been proposed. Unfortunately, each designer promoted his/her idea focused on specific application categories and with restricted sets of functionalities, lacking any sort of significant coordination effort. Thus, overlapping and not interoperable functionalities have often been produced. We recall project JXTA (launched by Sun MicroSystems) \cite{JXTA}, and the IETF P2P Working Groups: Peer-to-Peer SIP signaling (P2PSIP), Application-Layer Traffic Optimization (ALTO), and Low Extra Delay Background Transport (LEDBAT).

\subsubsection{Examples}
Very popular HM-based P2P systems are Soulseek \cite{Soulseek}, Napster \cite{Napster}, eMule \cite{eMule1} \cite{eMule2}, and BitTorrent \cite{BitTorrent}, which are mostly used for content sharing. BitTorrent is also used for live and on-demand streaming, such as in the UE-funded P2P-Next project (having BBC as early content partner) \cite{P2PNext}.

The Gnutella DUM-based protocol \cite{Gnutella} was released by Nullsoft in late 1999. Currently, it is at the 6th release. In Gnutella, the routing protocol is independent from the topology, which is random. Resources and their descriptors are kept locally by every peer. 
 
Distributed K-ary Search (DKS) \cite{Alima03} is the most popular DSM-based protocol, including Chord \cite{Stoica03} and Pastry \cite{Buyukkaya08}. Every DKS instance is a decentralized overlay network defined by the following parameters: the maximum number of nodes $N$, the search arity $k$, and the degree of fault tolerance $f$. 
The overlay network resulting from the instantiation of these parameters has several desirable properties. 
\textit{E.g.}, lookups are resolved in $log_kN$ overlay hops each, at most. 
Moreover, each node has to keep only $(k + 1) log_kN + 1$ addresses, in its routing table.
Chord \cite{Stoica03} is the most famous DKS($N, 2, f$) protocol. Given a key (\textit{i.e.} the identifier of a resource or a service), the protocol assigns the key to a node (a host or a process identified by an IP address and a port number). Chord's routing algorithm is illustrated in the subsection related to the \textbf{Message Routing} pattern (see below).

FastTrack \cite{FastTrack} is the LM-based protocol adopted by KaZaA \cite{Kazaa}, iMesh \cite{iMesh} and other popular file sharing applications. Moreover, FastTrack was used by early (pre-Microsoft) Skype \cite{Skype} releases. 
FastTrack is an extension of the Gnutella protocol, using supernodes that improve scalability and support the stability of the network, by acting as indexing servers.
Every peer maintans a list of active supernodes, for future connection needs.
In the file sharing context, leaf peers upload their file lists to their reference supernodes.
Search requests are sent to supernodes, as well. A search request that cannot be satisfied by a supernode is forwarded to other supernodes.
Once a partial or complete copy of the searched file has been located, provider and consumer peers start the data transfer process, which is based on the HTTP protocol. Despite the protocol that deals with leaf-supernode communication has been reverse-engineered, 
the protocol for supernode-supernode communication is still largely unknown.

Wuala \cite{Wuala} is a popular LM-based distributed storage systems, adopting a Chord-like \textbf{Overlay Scheme} for organizing supernodes, each one managing a set of storage nodes.

SCSP \cite{Liu2011} is an LM-based service provision framework, where supernodes play a key role in building the coordination among service groups (S-labor-market model). An experimental evaluation showed that SCSP is efficient, scalable and robust.

We complete this (non-exhaustive) survey by considering the JXTA platform \cite{JXTA}, which provides core building blocks (such as IDs, advertisements, peergroups and pipes) as well as a default set of core overlay policies, which can be (not-so-easily) replaced if necessary. The
overlay network is $2$-layered, with two primary node types:
\begin{itemize}
\item rendezvous super-peers (supernodes), which are connected according to a DUM-based scheme; super-peers are able to propagate queries and keep pointers to edge peers that cache resource advertisement;
\item edge peers (leaf nodes), which are able to send query/reply messages, not to propagate queries.
\end{itemize}
Every edge peer is connected to one and only one rendezvous super-peer. Every rendezvous super-peer has a
Rendezvous Peer View (RPV), \textit{i.e.}, an ordered list of known rendezvous super-peers.

\subsubsection{Consequences}
The adoption of an adequate \textbf{Overlay Scheme} guarantees effectiveness and efficiency of the global services offered by the P2P system.  

\subsubsection{Related Patterns}
\textbf{Topology}, \textbf{Message Routing}.

\subsection{\textbf{Bootstrapping}}

\subsubsection{Context} 
In any open peer-to-peer overlay network, the initial discovery of other peers participating in the network is one key operation. 

\subsubsection{Problem}
How to make incoming peers join the overlay network?

\subsubsection{Forces} 
\begin{enumerate}
\item Isolated peers need to join the network.
\item Joining peers should not connect all to the same peer --- conversely, peers' selection for join should either be random or driven by application-specific needs. 
\end{enumerate}

\subsubsection{Solution}
\begin{figure}[h!]
\centering
\includegraphics[width=10cm]{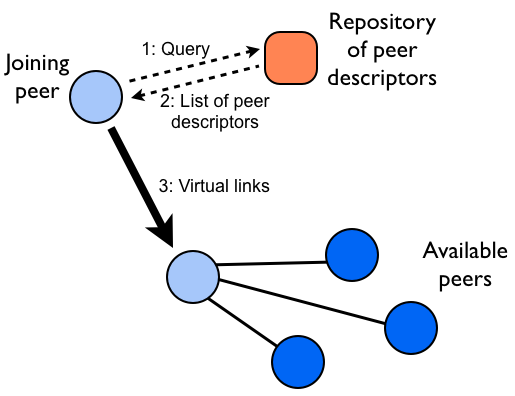}
\caption{\textbf{Bootstrapping} pattern.}
\label{fig:bootstrapping}
\end{figure}

Allow joining peers to obtain a list of peers which should be online, to let them virtually link to at least one peer that is already a member of the overlay network. If no such overlay network exists, the searching peer must form a new overlay network, which can be discovered and joined by further peers. The phases of the pattern are illustrated in Figure \ref{fig:bootstrapping}.

\subsubsection{Variations}
\begin{figure}[h!]
\centering
\includegraphics[width=10cm]{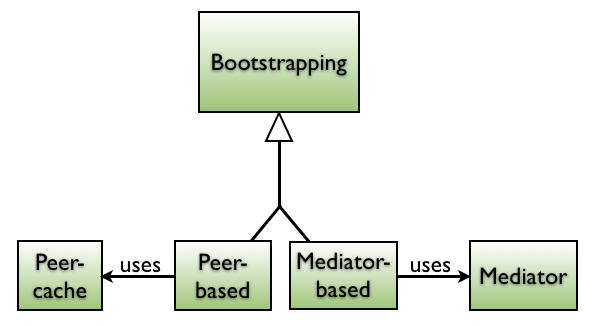}
\caption{Variations of the \textbf{Bootstrapping} pattern.}
\label{fig:varBootstrapping}
\end{figure}

The solution can be either \textit{peer-based} or \textit{mediator-based} (Figure \ref{fig:varBootstrapping}) \cite{Knoll08}. 

New peers may be discovered by asking known peers (peer-based approach). A \textit{peer-cache} is a list of previously known peers. When joining the overlay network, a peer tries to contact one of the peers in its peer-cache. If available, the contacted peer can be used as an entry point into the overlay network. Despite its simplicity and high efficiency, such an approach cannot guarantee that bootstrapping succeeds, as it may happen that contacted peers do not collaborate.

Conversely, mediator-based approaches adopt a well known entry point, \textit{i.e.}, the \textit{mediator}, to support the discovery process. For example, the mediator can be a server managing a list of online peers and pointing newly joining peers to one of them. In this approach, it is mandatory to keep the mediator's data updated and to make sure that the mediator is always available. As long as this is guaranteed (consuming a considerable amount of resources), the mediator-based approach enables bootstrapping. 

\subsubsection{Examples}
Early (pre-Microsoft) Skype releases, characterized by LM-based overlay, adopted the mediator-based bootstrapping approach. A login server was used to register first-coming peers, providing them with credentials (signed certificates) and ensuring their usernames to be unique. Every time a registered Skype peer wanted to join the network, it had to connect to the login server to authenticate itself and to receive a list of already connected superpeers. Such a list was frequently updated by the login server.

BitTorrent uses the mediator-based approach as well. A peer that wants to share a content item (file or directory), creates a .torrent static metainfo file and publishes it to one of the torrent Web servers. Every server is responsible for linking the .torrent file with a suitable tracker, a software agent that keeps a global registry of all providers of the corresponding file. To find a .torrent, a browser and a search engine are enough. Once the peer is connected to the suitable tracker, the latter one provides a list of peers that are currently downloading the content item.

The peer-based bootstrapping strategy is used for example in eMule, which has a HM-based overlay. In eMule, peers connect to the network for sharing files, and publish file descriptions to file indexing servers. Every peer has a cached list of servers, which can be easily updated.

\subsubsection{Consequences}
\textbf{Bootstrapping} prevents peers from being isolated and lets them join the overlay.  

\subsubsection{Related Patterns}
\textbf{Topology}.

\subsection{\textbf{Topology}}

\subsubsection{Context}
Peer-to-peer networks are highly dynamic and totally lack centralized control. Thus, their topologies are usually considered as emerging properties that may be predicted more or less accurately, according to the \textbf{Overlay Scheme}.

\subsubsection{Problem} 
How to model the topology of the overlay network? 

\subsubsection{Forces}
\begin{enumerate}
\item When designing a peer-to-peer network, it would be very useful to know in advance the features of its possible topologies, in order to predict (\textit{e.g.}, by means of simulations) the effectiveness of overlay routing strategies.
\item Topology awareness may be helpful also at runtime, in order to dynamically adapt parameters like the Time-To-Live (see the \textbf{Message Routing} pattern for details).
\item Topology construction mechanisms should be simple; in particular, it should not be necessary to flood the network with control messages.
\item Every peer should have a peerview (\textit{i.e.}, a constrained list of references to other peers) of adequate size. 
\end{enumerate}

\subsubsection{Solution} 
Describe the topology in terms of distribution of the node degree, clustering coefficient, average connected distance and diameter.

Refferring to the theory of random graphs, a network topology can be described in terms of \textit{node degree} distribution:
\begin{equation*}
  P(k) = P\{\textrm{\text{node degree}} = k\}
\end{equation*} 

The \textit{clustering coefficient} $CC$ is another important feature, defined as follows. Suppose that $i$-th node has $k_i$ neighbors. Then one may find at most $k_i(k_i - 1)/2$ links between such neighbors (namely, when they form a complete graph). Let $n_i$ be the actual number of links between the neighbors of the $i$-th node. Then, we define the clustering coefficient for the $i$-th node as
\begin{equation}
{CC}_i = \frac{n_i}{k_i(k_i - 1)/2}
\end{equation}
By averaging ${CC}_i$ over all nodes, the clustering coefficient for the whole network is obtained as
\begin{equation}
CC = \frac{1}{N} \sum_{i=1}^{N}{{CC}_i}
\end{equation}

Other important parameters for evaluating a topology are the \textit{average connected distance} $\langle l \rangle$ (also known as average path length, or separation degree), which is the expected value of the shortest path length between node pairs in a network, and the \textit{diameter} $d$, which is the maximal geodesic distance, considering all possible pairs of nodes.

\subsubsection{Variations}
\begin{figure}[h!]
\centering
\includegraphics[width=7cm]{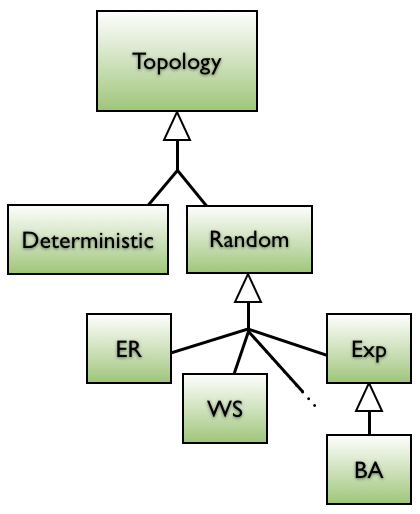}
\caption{Variations of the \textbf{Topology} pattern.}
\label{fig:topology}
\end{figure}

As shown in Figure \ref{fig:topology}, there are a many strategies, either deterministic or random, for describing and analyzing overlay network topologies. 

DSM-based P2P overlay network build their topologies according to precise strategies, for which they can be considered as ``deterministic'' topologies. Usually, all peers have a routing table with the same, constrained size. The node degree has a narrow Gaussian distribution.

DUM-based P2P overlay networks have purely random topologies, such as those based on the model introduced by Erd\"{o}s and R\'enyi (ER model)~\cite{Bollobas98}. ER topologies have $N$ vertices, each one having $\langle k \rangle = \alpha$ links on average. A link between two vertices may be present or not, independently of the presence or lack of any other link. Thus, a link is considered to have independent probability $p$. Trivially,
\begin{equation}
p = \frac{\alpha}{N-1}
\end{equation}
With independent nodes, the node degree has a binomial distribution:
\begin{equation}
P(k) = {N-1 \choose k} p^k (1-p)^{N-1-k}
\end{equation}
converging to the Poisson distribution for large $N$ values:
\begin{equation}
P(k) = \frac{\alpha^k e^{-\alpha}}{k!} \text{\hspace{0.7cm} with $\alpha = \langle k \rangle = {\sigma}^2$}
\end{equation}

ER networks have small clustering coefficient:
\begin{equation*}
{CC}_{ER} = \frac{2L}{N(N-1)} = \frac{\alpha}{N-1} = p
\end{equation*}
which is why the ER model is unsuitable for describing networks characterized by short average connected distance (\textit{small world phenomenon}). Like social networks, peer-to-peer networks have knowledge links between participants. Two peers know each others if any time one can open a communication channel with the other. Since the average connected distance reflects the expected time for transfering information from one node to another, the small world phenomenon is highly desirable in computer networks and in particular in peer-to-peer networks.

Watts e Strogatz~\cite{Watts98} introduced a simple model (denoted as WS model) for describing network topologies with ${CC}_{WS} \gg {CC}_{ER}$ and $\langle l \rangle_{WS} \gtrsim \langle l \rangle_{ER}$. In the WS model, a random rewiring procedure is used for interpolating between a regular ring lattice and a purely random topology, while preserving the number of links and nodes in the graph. In detail, the starting point is a ring lattice of $N$ nodes, each one being connected to $K/2$ other nodes on each side. Then, each link is rewired with probability $p_{WS}$, where rewiring means moving one end of the link to a new node, randomly chosen from the whole network, assuming that no two nodes can have more than one link and a node cannot be self-linked. If $p_{WS} = 0$, the lattice is highly ordered, with $\langle l \rangle_{WS} = \frac{N}{2K}$. Instead, if $p_{WS} \rightarrow 1$ the topology becomes a purely random one, with $\langle l \rangle_{WS} \rightarrow \frac{\ln{N}}{\ln{\alpha}}$.
In the case of the ring lattice, the node degree distribution equals $N$ at $K$, zero elsewhere. When $p_{WS} \rightarrow 1$, the node degree has a Poisson distribution. 

The WS model produces graphs with homogeneous node degree (\textit{egalitarian small world networks}). Several large real-world networks (such as the World Wide Web, scientists, food chains, rivers, top managers, movie actors) resulted to belong to a class of inhomogeneous networks, characterized by node degree distribution decaying as a \textit{power law}, \textit{i.e.}, a polynomial relationship with the property of scale invariance:

\begin{equation}
P(k) = c k^{-\tau}
\label{eq:powerlaw}
\end{equation}

with $\tau > 1$ and $c$ normalization factor. Several studies have proven that such \textit{scale-free} networks are highly robust, \textit{i.e.}, local failures rarely deteriorate the global ability of the network to carry information. 

Barab\'asi and Albert~\cite{Barabasi99A} proposed a generative model (known as BA model) that builds scale-free networks with $\tau \simeq 3$. More precisely, the resulting node degree distribution is
\begin{equation}
P(k) = \frac{2m(m+1)}{k(k+1)(k+2)} \simeq 2m^2k^{-3} \quad \forall k > m
\label{eq:ba}
\end{equation}

The BA model is based on growth and preferential attachment. A network that grows without preferential attachment, \textit{i.e.}, with every new peer choosing $m$ existing peers ($m \in [1,N_0]$ uniformly random) with even probability, results in having an exponential node degree distribution:
\begin{equation}
P(k) = (1-e^{- \frac{1}{m}})e^{1-\frac{k}{m}} \quad \forall k \geq m
\end{equation}

The P2P designer should be aware that, in terms of robustness against random failures, scale-free are better than any other type of topologies. The reason is simple: the probability that most connected peers fail is low, due to their scarcity. On the contrary, scale-free networks are not robust against targeted attacks \cite{Albert2000} --- indeed, disconnecting a hub would result in breaking the network in a number of isolated clusters.

\subsubsection{Examples}
In Chord \cite{Stoica03}, which is a DSM-based system, a node $n$ starts its activity by calling $n.join(n')$. Here, $n'$ is any known Chord node, and the $join()$ function forces $n'$ to find the successor of $n$, \textit{i.e.} the first peer whose identifier is greater than the identifier of $n$.  
For example, suppose node $n$ joins the system, with an ID lying between nodes $n_p$ and $n_s$. By means of the $join()$ function, $n$ sets $n_s$ as its successor. When notified by $n$, node $n_s$ sets $n$ as its predecessor. Node $n_s$ is then asked for its predecessor (which is now $n$), when $n_p$ runs $stabilize()$. In this way, $n_p$ sets $n$ as its successor. Finally, $n_p$ notifies $n$, and $n$ sets $n_p$ as its predecessor. The resulting topology is an almost deterministic, highly regular topology, with node degree distribution looking like a Gaussian function centered in $m/2$, as every node keeps a routing table filled with up to $m$ entries (more details are provided in the description of the \textbf{Message Routing} pattern). 

Ripeanu showed that despite Gnutella (a DUM-based system) is not a pure power-law network, its structure has the advantages and drawbacks of a power-law one, and does not fit well the underlying Internet topology, thus leading to ineffectual use of the physical infrastructure \cite{Ripeanu99}.

According to Kleinberg \cite{Kleinberg99}, Freenet (another DUM-based system) should retrieve data items in about $O(log_2N)$ hops, if peers are linked in a small-world network. This is possible on mature Freenet networks only, and may be prevented by free riders that produce high churn rate. 

\subsubsection{Consequences}
Topology awareness is useful both during the design phase of the P2P network, and also at run-time, to predict/analyze its performance as well as to adjust functional parameters.  

\subsubsection{Related Patterns}
\textbf{Message Routing}, \textbf{Distributed State}, \textbf{Group Membership}.

\subsection{\textbf{Message Routing}}

\subsubsection{Context}
Message propagation among peers, based on distributed mechanisms, is one of the keys of the correct operation of a P2P system. 

\subsubsection{Problem} 
How are messages propagated in the overlay network?
It would be unusual that one peer acts as gateway between any couple of other peers of the overlay network. Instead, every peer, with its limited knowledge of the network (\textit{peerview}), should contribute to the process of forwarding messages, even if the peer has no interest in the content of the message (which may be encrypted).  

\subsubsection{Forces}
\begin{enumerate}
\item Every peer must be able to propagate messages to some or all of its neighbors (\textit{i.e.}, members of its peerview).
\item Propagation may be random or dictated by a global plan, whose benefits emerge from full cooperation among peers.
\item According to the \textbf{Overlay Scheme}, the global routing strategy may take into account the network topology, or not.
\end{enumerate}

\subsubsection{Solution} 
\begin{figure}[h!]
\centering
\includegraphics[width=10cm]{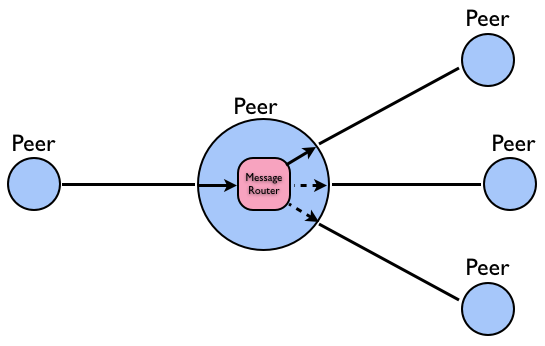}
\caption{\textbf{Message Routing} pattern.}
\label{fig:MessageRouting}
\end{figure}

Define a component that consumes messages from a message channel. Consumed messages are forwarded each one to a different message channel, depending on a set of conditions. The solution is illustrated in Figure \ref{fig:MessageRouting}.
 
The \textbf{Message Routing} pattern uses the \textbf{Distributed State} pattern and proposes general mechanisms to propagate messages from one peer another (or many others) \cite{Hohpe03}.

\subsubsection{Variations}
In DUM-based systems, requests are usually propagated by flooding the network with messages. 
Such a strategy is efficient in small communities like company networks, but consumes much bandwidth, thus preventing scalability.
To mitigate this issue, recent research requests can be cached and probabilistic flooding can be used \cite{Sasson2003}. 
Moreover, by giving unique identities to messages, repeated occurrences of the same message can be deleted and loops can be stopped
from forming. Finally, it may be necessary to include a Time To Live (TTL) counter to put a constraint on the number of message propagations.

In DSM-based systems, the recipient is always selected according to rules that take into account the topology. 
Indeed, topology and routing strategy are strictly interrelated. 
Every peer has a randomly generated ID, belonging to the same key space adopted for resources. 
Every peer is responsible for caching $(key, value)$ pairs, for a small subset of the entire key space.
When a resource has to be advertised, the associated $(key, value)$ pair is routed towards the peer whose ID is most similar to the resource key. Such a process is repeated until the nearest key is the one of the current peer. Similarly, a resource query is forwarded towards the peer whose ID is most similar to the one of the resource.
We recall that in DSM-based \textbf{Overlay Scheme}s, the responsibility of storing knowledge about shared resources is much more evenly distributed among peers than in DUM-based \textbf{Overlay Scheme}s. 

\subsubsection{Examples}
\begin{figure}[h!]
\centering
\includegraphics[width=12cm]{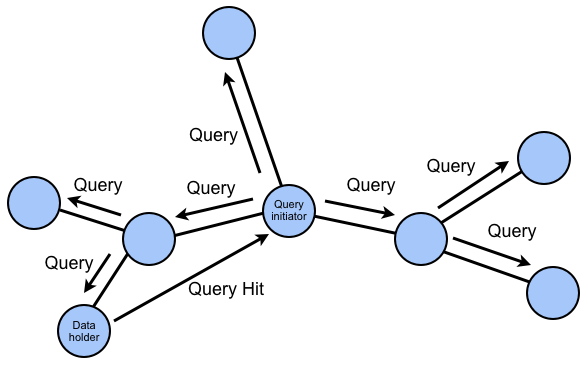}
\caption{The \textbf{Message Routing} strategy of Gnutella.}
\label{fig:gnutellaRouting}
\end{figure}

In Gnutella \cite{Gnutella}, routed messages are PING and QUERY, respectively used for peer and resource discovery. When received by a node, messages are relayed to all the peers in the node's peerview (such a flooding approach is illustrated in Figure \ref{fig:gnutellaRouting}). To mitigate network congestion, PING and QUERY messages are always associated to a TTL. PONG and QUERY HIT descriptors may only be sent along the same path of the related PING and QUERY messages. Once a resource provider and the consumer are in touch, their interaction is direct, unmediated.

Another DUM-based system that adopts message broadcasting, denoted as MUTE \cite{MUTE}, uses Utility Counters (UCs) instead of TTLs. UCs' purpose is to limit how far a query may travel, based on both the branching factor at each query hop (\textit{i.e.}, the number of neighbor a message is forwarded to) and the number of results the query generated. 

\begin{figure}[h!]
\centering
\includegraphics[width=12cm]{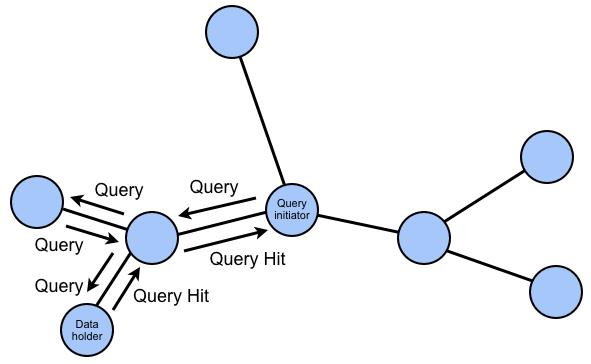}
\caption{In this Freenet example, a file request is routed through the network, finds itself in a dead-end and finally reaches the file owner.}
\label{fig:FreenetRouting}
\end{figure}

In Freenet \cite{Freenet} (which is DUM-based, too), every node has a routing table for associating node addresses with files that are supposed to be held by the considered nodes. Freenet nodes forward file lookup messages or file placement messages (Figure \ref{fig:FreenetRouting}).
Files with similar keys are clustered on the same node, if possible. 
Freenet uses a particular message and file forwarding scheme to ensure that neither the original requestor nor the actual file owner can be tracked.
Upon receiving a file request, a Freenet node first checks its own repository, and if finds the file, returns it to the node the request arrived from. The file is sent upstream, without revealing the node that owns it.
Otherwise, the node forwards the file request to the node that most probably owns the requested file, according to the routing table. That node then
checks its repository, and so on. Depending on the distance from the file owner, nodes along the path might also cache copies in their repositories.

\begin{figure}[h!]
\centering
\includegraphics[scale=0.5]{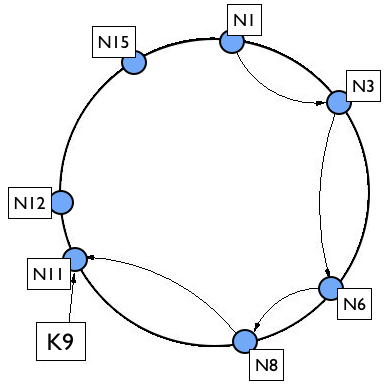}
\caption{Key search in a Chord ring, when the basic lookup algorithm is used.}
\label{fig:Chord_simple_lookup}
\end{figure}

In the DSM-based system called Chord, given a resource descriptor, the protocol maps it onto a node. 
Chord adopts consistent hashing to assign each node and resource an $m$-bit identifier (also dented as key). 
A node identifier is randomly generated, while a resource key is produced by hashing the resource descriptor. 
The identifier length $m$ must be large enough to make it unlikely that two different nodes or resources hash to the same identifier. 
A published resource descriptor with key $k$ is forwarded until it reaches the node whose identifier equals $k$ or the first node whose identifier follows $k$.
Such a node is denoted as the successor node of key $k$. 
According to the Chord basic lookup algorithm, key lookup works if and only if every node is able to communicate with its current successor node on the Chord ring. By means of successor pointers, queries for a given key are passed around the ring, until they find the node that stores the sought resource descriptor. Figure~\ref{fig:Chord_simple_lookup} shows an example in which node $N1$ performs a query for key $K9$. 

\begin{figure}[h!]
\centering
\includegraphics[scale=0.5]{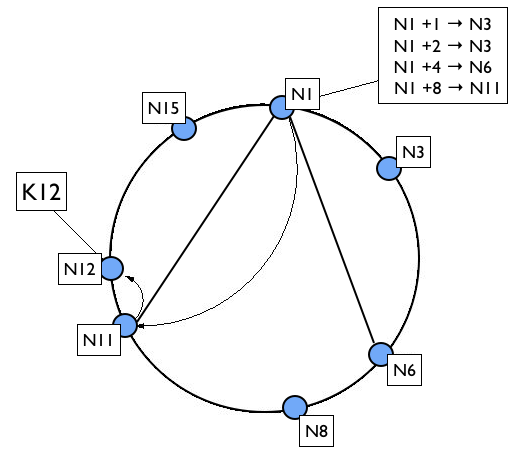}
\caption{Key search in a Chord ring, when the scalable lookup algorithm is used.}
\label{fig:Chord_fingered_lookup}
\end{figure}

Chord's basic lookup algorithm uses a number of messages that is linear in the number of nodes. 
With Chord's scalable lookup algorithm (illustrated in Figure \ref{fig:Chord_fingered_lookup}), the search process is accelerated by means of the routing table every node maintains, denoted as \textit{finger table}.
The $i^{th}$ entry in the finger table at node $n$ contains the key, IP address and port of the first node $s$ that follows $n$ by at least $2^{i-1}$ in the Chord ring. Formally, $s = successor(n+2^{i-1})$, where $1 \leq i \leq m$ (it is worth noting that all the arithmetic is modulo $2^m$). Node $s$ is denoted as the $i^{th}$ \textit{finger} of node $n$, namely $n.finger[i]$. Finger allows peer for skipping a large section of the Chord ring, in the lookup process. It has been proved that the number of nodes that have to be involved when looking for a key successor in an $N$-node network is $O(\log N)$, with high probability.

\subsubsection{Consequences}
When carefully designed, \textbf{Message Routing} is the cornerstone of the P2P system.

\subsubsection{Related Patterns}
\textbf{Overlay Scheme}, \textbf{Distributed State}.

\subsection{\textbf{Distributed State}}

\subsubsection{Context}
Peer-to-peer systems work correctly, efficiently, and robustly, if peers share their resources.

\subsubsection{Problem} 
How to name, put, replicate and get resources within the overlay network?

\subsubsection{Forces}
\begin{enumerate}
\item No two different peers or resources can share the same identifier.
\item The assignment of identifiers should not be centralized, unless required by severe security restrictions.
\item Peer and resource descriptors should have a ``standard'', machine-readable format.
\item Since a discovered resource may be a replica, its descriptor must contain a precise reference to the actual owner of that replica.  
\end{enumerate}

\subsubsection{Solution} 
Let each resource have a unique identifier, and provide peers with PUT, STORE, GET, LOOKUP mechanisms.

\begin{figure}[h!]
\centering
\includegraphics[width=10cm]{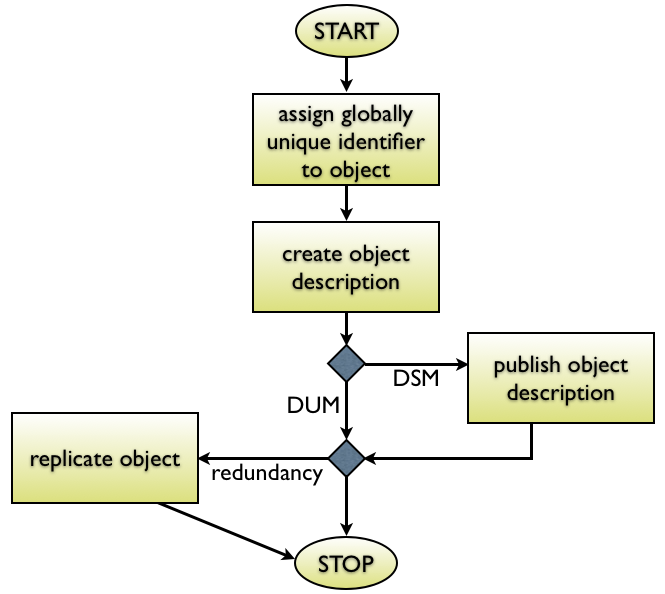}
\caption{Flowchart illustrating the main operations enabled by the Distributed State pattern.}
\label{fig:DistributedState}
\end{figure}

Peers and shared resources must be characterized by unique identifiers, in order to be advertised and discovered within the overlay network. 
The \textbf{Distributed State} pattern (Figure \ref{fig:DistributedState}) builds upon a couple of existing patterns for object identification and redundancy (namely, the \textbf{Object Identifier} and the \textbf{Redundant Independent Objects}), providing a robust mechanism for identifying peers and resources within the network. 

The \textbf{Object Identifier} pattern allows one to assign a globally unique identifier to objects (thus, to peers and shared resources), granting them a unique identity within the overlay network \cite{Hohpe03}. 

The \textbf{Redundant Independent Objects} pattern ensures the availability of a shared resource even if the one or more host peers experience a failure. This is achieved by replicating the shared resource on several different hosts \cite{Hohpe03}.

Moreover, mechanisms for placing (PUT, STORE) and locating (GET, LOOKUP) resources, but also peers (PING), can be implemented using the \textbf{Lookup} pattern \cite{Kircher00}, which describes how to find and retrieve initial references to distributed objects and services. 

In DUM-based P2P overlay networks, \textbf{Lookup} is only used for locating peers and resources, as no advertising is performed. In DSM-based overlay networks, publication and search usually adopt the same strategy, reason why the \textbf{Lookup} pattern is used in both cases. 

In HM-based P2P systems, the \textbf{Registry} pattern is sufficient, providing a service that takes a resource name and returns a descriptor that encapsulates the knowledge of how to get the resource \cite{Hohpe03}.

\subsubsection{Examples}
The examples we proposed for the \textbf{Message Routing} pattern also apply to \textbf{Distributed State}. 

Until the introduction of magnet links, there was no standard solution for global unique identifiers and peer/resource descriptors, \textit{i.e.}, each P2P system had its own. A good approach, in our opinion, was the one proposed by the JXTA specification. There, an ID Format is a scheme for identifiers of JXTA entities. Each ID Format is associated to a sub-namespace of the ``jxta'' URN namespace. An ID Type describes the features of JXTA identifiers that refer to a particular kind of JXTA entity, currently including peergroups, peers, codats, pipes, module specifications, module classes and module implementations. JXTA Advertisements are XML documents describing peers, peergroups and services. 

MAGNET \cite{MAGNET} has emerged as  an open, vendor- and project-neutral URI scheme that defines the format of magnet links. MAGNET is currently a de facto standard for content-based identification of a file through a cryptographic hash value, irrespective of its location. The suitability to P2P systems lies in the fact that magnet links allow resources for being referred to without the need for an always available host, and can be easily generated by any peer who already owns the file, thus there is no need for a central issuer.

Secure Hash Algorithm (SHA) \cite{SHA} is a family of cryptographic hash functions published by the National Institute of Standards and Technology (NIST) as a U.S. Federal Information Processing Standard (FIPS).
In general, a SHA hash may be used as both a global unique ID and integrity check for files. For example, magnet links use SHA-1 for both purposes.

\subsubsection{Consequences}
\textbf{Distributed State} allows peers for managing shared resources --- their own ones, or those provided by other peers.

\subsubsection{Related Patterns}
\textbf{Data Protection}, \textbf{Choke/Unchoke}, \textbf{Multisource Data Transfer}.

\subsection{\textbf{Information Consistency}}

\subsubsection{Context}
Resource sharing is effective only if related information is consistent.

\subsubsection{Problem} 
How to preserve the consistency of resource descriptors within the overlay network?
Peer-to-peer systems adopt distributed strategies for sharing resources, abstaining from centralized control. To this purpose, peers also need to store information about shared resources --- \textit{e.g.}, resource descriptors. Information lookup is a general feature of all \textbf{Overlay Scheme}s, while  only HM-based and DSM-based \textbf{Overlay Scheme}s provide mechanisms for remote information publishing. In DUM-based systems, information about owned resources is kept locally. The advantage of remote publishing is that information about the existence of a resource can be replicated on several peers, thus facilitating the lookup process. On the other hand, remote publishing may produce inconsistencies, if the resource description is changed by the resource owner and republished. 

\subsubsection{Forces}
\begin{enumerate}
\item Peers need to find resources within the overlay network, for which replicating each resource descriptor on multiple nodes would help.
\item Too many copies of a resource descriptor spread over the network could become a problem if their information is obsolete.
\end{enumerate}

\subsubsection{Solution} 
\begin{figure}[h!]
\centering
\includegraphics[width=10cm]{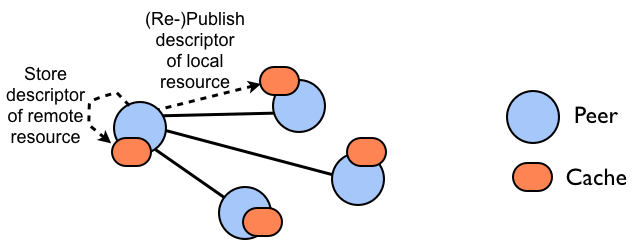}
\caption{The \textbf{Information Consistency} pattern.}
\label{fig:InformationConsistency}
\end{figure}

For resource owners, let them choose the most convenient place to store resource descriptors, assign limited lifetimes to such descriptors, and periodically refresh them --- \textit{i.e.}, republish them to the same cache, or to a more convenient one that may have appeared meanwhile. 
For requesters, if a discovered resource is something that can be reused (like a service), let them keep track of the resource owner. 
These two aspects of the same solution are shown in Figure \ref{fig:InformationConsistency}.
 
\subsubsection{Variations}
\begin{figure}[h!]
\centering
\includegraphics[width=10cm]{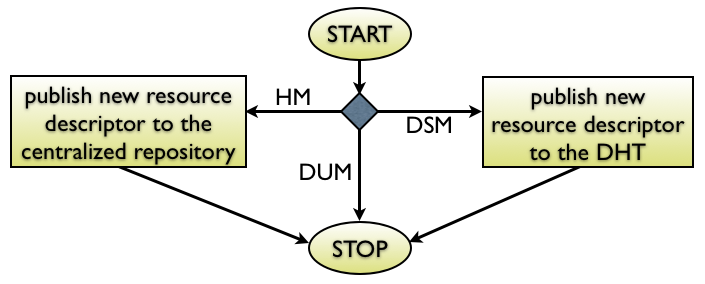}
\caption{Flowchart illustrating the three variations of the \textbf{Information Consistency} pattern.}
\label{fig:varInformationConsistency}
\end{figure}

Depending on the \textbf{Overlay Scheme}, different approaches are used (Figure \ref{fig:varInformationConsistency}).
\begin{itemize}
\item In HM-based systems, a centralized repository is used for storing resource descriptors. Thus, when a resource descriptor gets updated, the resource owner must notify the centralized repository.
\item DUM-based systems adopt the ``pull'' approach, reason why resource owners do not publish resource descriptors. Peers that propagate requests may be traversed by responses, for which they may cache the descriptors of discovered resources. These documents may be replaced afterwards, if updated copies pass through the peer, or may be deleted, if they have an expiration time.
\item DSM-based systems use the ``push'' approach, where resource descriptors are published to specific peers, with uniform responsibility share-out. If the network were stable, it would be simple, for a resource owner, to send updated resource descriptors to responsible nodes. Unfortunately, network dynamics result in responsibility redistribution, which is why it would be unreasonable to implement a messaging protocol for keeping the resource owner aware of which peers store its resource descriptors. Actually, in most DSM-based \textbf{Overlay Scheme}s, resource descriptors expire after a limited time interval, whose length depends on the application.  
\end{itemize}

\subsubsection{Examples}
In eMule \cite{eMule1}, which is a HM-based system, in case a peer has files to offer, the ``offer-files'' message is sent as soon as the connection with the index server has been established. The message is also sent when the list of shared files is updated. The ``offer-files'' message contains the number of files described within, and the list of file descriptors \cite{eMule1} --- anyway, no more than $200$ (the server can also set a lower bound to this number).

Unlike other DUM-based P2P file sharing systems, such as Gnutella \cite{Gnutella}, Freenet \cite{Freenet} stores replicates data items along downstream routes, acting as a large distributed cache. To this purpose, every node allocates some quantity of disk space for storing data items. A user interested in sharing a file ``inserts'' it into the network. When the ``insertion'' process is complete, the publisher can turn off her/his peer, as the file is stored within the network and will stay online. This also implies that no file descriptor has to be published into the network. Two advantages of this approach are anonymity and high reliability. Moreover, Freenet is unaffected by one typical BitTorrent problem, \textit{i.e.}, the lack of nodes (denoted as ``seeds'') that provide full replicas of a file. The main drawback of the storage method is that Freenet ``forgets'' data which are not frequently retrieved. Moreover, while inserting data into the network is always possible, the immediate cancellation of a file cannot be forced. 

Finally, let us consider Chord \cite{Stoica03}, which is a DSM-based \textbf{Overlay Scheme}. Malicious or buggy Chord participants could provide an incorrect view of the Chord ring (we described it within the description of the \textbf{Message Routing} pattern). One way to check global consistency is to periodically ask other nodes to perform a lookup for the key of the requesting node itself. If the lookup process does fails, this could be an indication involved nodes have not a globally consistent view of the Chord ring. A similar strategy is used to periodically discover which peers are responsible for storing resource descriptors. Of course, to prevent such a strategy from being too much expensive in terms of bandwidth consumption, the lookup period must be reasonable.

\subsubsection{Consequences}
\textbf{Information Consistency} supports the discovery of resources which are actually available within the overlay network.

\subsubsection{Related Patterns}
\textbf{Distributed State}.

\subsection{\textbf{Data Protection}}

\subsubsection{Context}
Resource sharing is effective if related data are protected against corruption and if it is possible to discriminate between correct copies and corrupted ones.

\subsubsection{Problem}
How to guarantee the integrity, authenticity and availability of data stored in the P2P system?
Information sharing is easier when users can freely join and leave the system without any additional infrastructure to identify nodes. However, an open system increases the difficulty of enforcing the ``protection'' aspect. Indeed, the possibility to authenticate peers simplifies the process of integrating verification, ensuring that its content is authentic, that it is available, and that it has not been altered or compromised during transmission.

\subsubsection{Forces}
\begin{enumerate}
\item A peer that requests a document should receive only authentic responses from peers that are genuine.
\item A peer that requests a document should always find it available even if peers may leave the network.
\item A peer that requests a document should receive it uncorrupted and complete.
\end{enumerate}

\subsubsection{Solution} 
\begin{figure}[h!]
\centering
\includegraphics[width=10cm]{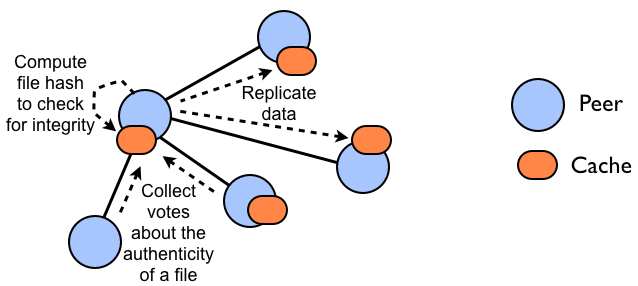}
\caption{The \textbf{Data Protection} pattern.}
\label{fig:DataProtection}
\end{figure}

Introduce a voting mechanism, sign data, replicate data either actively or passively, adopt secure routing policies in order to prevent malicious nodes polluting forwarded data (as illustrated in Figure \ref{fig:DataProtection}).

\subsubsection{Variations}
Several definitions of authentic file have been introduced and, therefore, different approaches have been proposed. 
The most reasonable definitions are \textit{expert-based}, \textit{voting-based}, \textit{reputation-based} and \textit{oldest document}.
In systems using an expert-based definition, an authoritative node stores the signatures for all its users' documents and decides which documents are authentic and which are not. Such a solution is weak, because the authoritative node is a single point of failure for the entire system. Voting-based systems mitigate this issue by involving several expert nodes for authenticating files, on a majority basis. Again, such an approach is easy to break, by means of spoofing techniques (applied to votes, nodes, and files). To improve the voting-based strategy, node reputation can be used a a weight for votes~\cite{Daswani2002}. The \textbf{Reputation} pattern is illustrated below.
Last but not least, in the oldest document definition, only the first sample of the document introduced in the system is assumed to be authentic. Clearly, this is the most restrictive definition of authentic file.
 
Data replication is a common way to ensure file availability, regardless of which nodes in the network are currently online and of file popularity.
The basic replication technique is denoted as \textit{passive replication}, naturally happening in file sharing systems, as peers continuously request and copy files from one another. \textit{Active replication} approaches include the forced migration of content to improve data locality and the caching of data items while they are routed throughout the network. Advanced forms of replication try to dynamically minimize the number of copies, while meeting quality of service constraints~\cite{Androutsellis2004}.

Finally, pursuing data integrity requires avoiding corruption of the data for communication failures. The file integrity problem is usually addressed by including message redundancy in the form of a signature. Common techniques are cyclic redundancy checks (CRCs), hashing, message authentication codes (MACs) and digital signatures. Additionally, secure routing can be adopted, in order to avoid malicious nodes to pollute the transferred data items.

\subsubsection{Examples}
Most P2P systems do not provide guarantees about data availability. In systems like Gnutella~\cite{Gnutella} and Freenet~\cite{Freenet}, files can disappear when they are no more requested.

Fortunately, the implementation of replication mechanisms is not too difficult. Freenet itself, for instance, applies cache-based file replication, during routing processes (upstream and downstream). More advanced replication techniques are employed by OceanStore~\cite{Rhea2003}, where data are migrated to areas of use. Replication is still a challenge for systems that correlate object identifiers to nodes, such as in Chord~\cite{Stoica03}.

Content sharing systems like eMule~\cite{eMule1} and BitTorrent~\cite{BitTorrent} use file hashing to quickly compare files that are claimed to be copies of the same original version. Suppose to be searching for a file and to find $m_1$ copies with the same hash, and $m_2 \ll m_1$ copies with another hash. With high probability, the first version (being much more replicated than the second one) is valuable, while the other version may be corrupted, incomplete, lower quality (\textit{e.g.}, if the file is audio/video) or absolutely wrong.

Skype~\cite{Skype} encrypts data chunks that are sent by peers, in order to make packet sniffing ineffective. 

\subsubsection{Consequences}
\textbf{Data Protection} preserves data against corruption and enables the discrimination between correct replicas and corrupted ones.

\subsection{\textbf{Group Membership}}

\subsubsection{Context}
To improve the efficiency and robustness of the system, peers with common interests and capabilities may join forces. 

\subsubsection{Problem}
How to avoid issues due to open or loosely controlled membership?

\subsubsection{Forces}
\begin{enumerate}
\item It should be possible to create interest-based peergroups, in order to scope query message propagation, thus improving lookup performance.
\item It is important to guarantee that message propagation remains within its scoped group.  
\end{enumerate}

\subsubsection{Solution} 
\begin{figure}[h!]
\centering
\includegraphics[width=10cm]{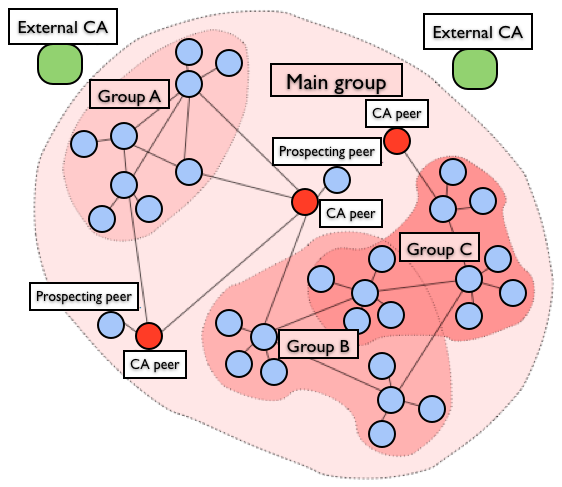}
\caption{Example of P2P system where the \textbf{Group Membership} pattern is used. Group access is based on certificates, which may be issued by internal or external Certification Authorities (CAs).}
\label{fig:GroupMembership}
\end{figure}

Create groups of peers with common interests and capabilities (Figure \ref{fig:GroupMembership}).

\subsubsection{Variations}
Despite being considered one of the basic functionalities for P2P systems, group management is still at an early stage of integration. Several solutions have been proposed, but there is no best one. 

Given a group of peers, message routing can be restricted to that group by tagging messages conveniently. Another approach is using a separate overlay for each group, \textit{i.e.}, applying a layered \textbf{Overlay Scheme} (LM-based)~\cite{Kassinen2009}.  

Regarding group organization, there are two alternative approaches: self-organization and user-driven subgrouping. 
With self-organization, peers spontaneously create coalitions, based on common interests~\cite{Castano2003}.
User-driven subgrouping is less challenging and more common.
 
Different group management styles have also been proposed. Elser \textit{et al.}~\cite{Elser2010}, for example, considered three group styles, characterized by different levels of security and efficiency. The ``Paradise'' style assumes the absence of malicious users and very weak anonymity. In the ``Monarchy'' style, a group of leader creates and fully control the groups. In the ``Voting'' style, selected members vote to accept or reject join requests from prospecting peers.

\subsubsection{Examples}
BitTorrent's peergroup management mechanism is widely known. User-driven peers interested in downloading or seeding a specific file are grouped together by a tracker, \textit{i.e.}, a software agent associated to the file since its first publication. Precisely, BitTorrent's peergroups are denoted as torrents.

A decentralized strategy for setting up peergroups is available in JXTA.  Every peer is a member of a global peergroup (namely, the NetPeerGroup) and is allowed to join one or more subgroups. As JXTA does not provide details about the creation or management of a group, researchers have proposed possible implementations~\cite{Amoretti2005b}.

An interesting attempt to develop a group management standard for P2P networks is the protocol proposed by Kassinen \textit{et al.}~\cite{Kassinen2009}.

\subsubsection{Consequences}
\textbf{Group Membership} allows peers for joining forces, to improve system efficiency and robustness.

\subsubsection{Related Patterns}
\textbf{Information Consistency}, \textbf{Reputation}.

\subsection{\textbf{Multisource Data Transfer}}

\subsubsection{Context}
It happens that several peers offer the same resource, while several other peers request it. While downloading the resource, a peer becomes a new potential provider of that resource.

\subsubsection{Problem} 
How to improve the performance of data transfer among peers?
Over the past decades, several techniques have been explored to improve data transfer rate and efficiency. 
A first possible obstacle is bandwidth limitation of receivers. Another possible barrier is the inability of sources to saturate the bandwidth of receivers. Congestion or failures at the network core may also slow the transfer. These issues impact P2P overlays as well.

\subsubsection{Forces}
\begin{enumerate}
\item A peer that needs some data pieces owned by another peer should be allowed to download them.
\item A peer that provides highly requested data pieces should not be the unique provider, in particular if its upload bandwidth is limited.
\end{enumerate}

\subsubsection{Solution} 
\begin{figure}[h!]
\centering
\includegraphics[width=10cm]{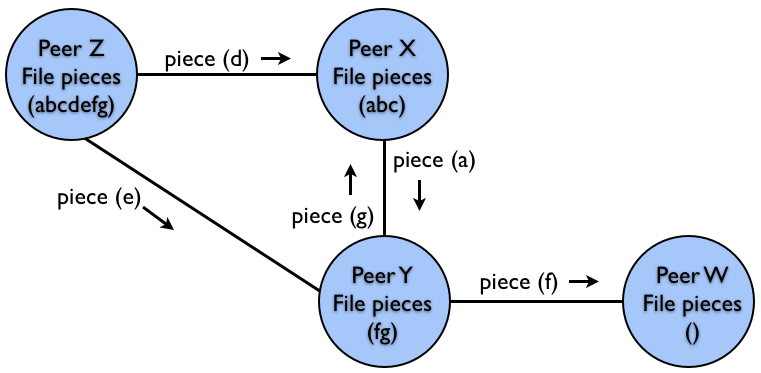}
\caption{Example of \textbf{Multisource Data Transfer}.}
\label{fig:mdt}
\end{figure}

Simultaneously download data from multiple sources (Figure \ref{fig:mdt}). In peer-to-peer overlays, this is a popular technique to accelerate transfers, when the receiver is not the bottleneck. 

Frequently, the \textbf{Multiple Data Transfer} pattern is completed by a strategy for selecting which data piece to download first, such as the Rarest First Algorithm (RFA) \cite{Legout06}. In RFA, \textit{available pieces} are those that have been already served at least once by the initial seed. The \textit{rarest pieces} have the least number of replicas in the peergroup. If the least replicated piece in the peergroup has $m$ replicas, then all the pieces with $m$ copies form the \textit{rarest pieces set}. Every peer has a periodically updated list of available replicas of each piece in its peergroup. Such a list is used to define a \textit{rarest pieces set}. Being $m$ the number of replicas of the rarest piece, the index of each piece with $m$ copies in the peergroup is added to the rarest pieces set. When a piece replica is added to or removed from the peergroup, the rarest pieces set is updated. The next piece to download is randomly selected in the rarest pieces set.

\subsubsection{Variations}
There are two basic strategies to locate data sources: \textit{per-file} and \textit{per-chunk} \cite{Pucha07}.
In a per-file system, downloaders locate other sources of the \textit{exact} file by means of $O(1)$ lookups, \textit{i.e.}, using a centralized indexer. Unfortunately, file transfers based on this approach are often intolerably slow (requiring hours or even days). 
In a per-chunk system, downloaders find sources for individual file pieces (chunks). Since any given file chunk might be replicated in several nodes, not necessarily keeping the whole file, this approach enables improved download performance. However, the cost is $O(N)$ lookups, one for each of the $N$ file chunks the downloader is trying to retrieve. 

\subsubsection{Examples}
BitTorrent~\cite{BitTorrent}, Gnutellat~\cite{Gnutella}, ChunkCast~\cite{Chun2006} are examples of per-file systems. BitTorrent uses RFA for piece selection.
Examples of per-chunk systems are Shark~\cite{Shark} and eMule~\cite{eMule1}, which is an implementation of the eDonkey2000 protocol, based on the MFTP multisource data transfer protocol.

In the P2P Live Streaming domain, systems based on the per-chunk approach have been deployed, supporting hundreds thousands of simultaneous users listening to the same channel, with high stream rates. Examples are CoolStreaming~\cite{Li08}, PPLive~\cite{Spoto09}, PPStream~\cite{Liang09}, UUSee~\cite{Liu2010}, and many more. Most of them adopt single-layer video. Instead, LayerP2P~\cite{Liu09} uses layered video encoding with nested dependency, where the video generated by lower layers is refined by a higher layer, and received multiple layers provide better video quality.

\subsubsection{Consequences}
\textbf{Multisource Data Transfer} allows to improve the performance of data transfer among peers.

\subsection{\textbf{Choke/Unchoke}}

\subsubsection{Context}
Voluntary resource sharing among peers is the basic premise of P2P systems. Nevertheless, the inherent tension between individual rationality and collective welfare threatens their viability~\cite{Feldman2005}. 

\subsubsection{Problem} 
How to guarantee a reasonable degree of data upload and download mutuality among peers?
Assuming that a specified protocol is blindly obeyed by all peers, in P2P networks where individual participants may interact with varying degrees of collaboration and competition, is not realistic. 
In P2P systems, cooperation may bump into significant computation and communication costs. Thus, rational peers (\textit{i.e.}, peers that try to maximize their own utility) may refuse to share their resources~\cite{Feldman2005}. 
Consuming resources from other peers without offering a \textit{quid pro quo} are denote as \textit{free-riders}. 
In 2000, Adar and Huberman~\cite{Adar2000} performed an extensive analysis of user traffic on Gnutella, establishing that almost 70\% of users shared no files at all. Five years later, Hugues \textit{et al.}~\cite{Hugues2005} performed a new analysis, showing an increasing downgrade of the network performance.
A robust \textbf{Multiple Data Transfer} protocol should be supported by incentive mechanisms for guaranteeing fair upload and download mutuality, while penalizing free riders. So far, only a few effective solutions that scale without central servers have been demonstrated.

\subsubsection{Forces}
\begin{enumerate} 
\item A peer that does not share downloaded data pieces should be penalized in subsequent download attempts.
\item Peers with higher download rates should be favored, with respect to slower peers.
\end{enumerate}

\subsubsection{Solution} 
Adopt a peer selection strategy that penalizes free riders, \textit{i.e.}, peers that download but never upload data.

We say that peer B is \textit{interested} in peer A when peer A has data pieces that peer B does not own. We also say that peer A \textit{chokes} peer B when peer A decides not to provide peer B with data. Conversely, peer A \textit{unchokes} peer B when peer A decides to provide peer B with data. Requests are pipelined, in order to avoid delay between data pieces being transferred, as when a sub-piece arrives a new request is served.
Peer A unchokes at most $M$ interested peers, using the following policy:
\begin{itemize}
\item Every $T_1$ seconds, interested remote peers are ordered with respect to their download rate, and the $M-K$ fastest peers are unchoked.
\item Every $T_2$ seconds, $K$ additional interested remote peers are unchoked at random. This random unchoke has two purposes. First, it does enable the download capacity evaluation of new peers. Second, it allows new peers that do not have any piece to share to bootstrap, by providing them their first piece.
\end{itemize}

\begin{figure}[h!]
\centering
\includegraphics[width=8cm]{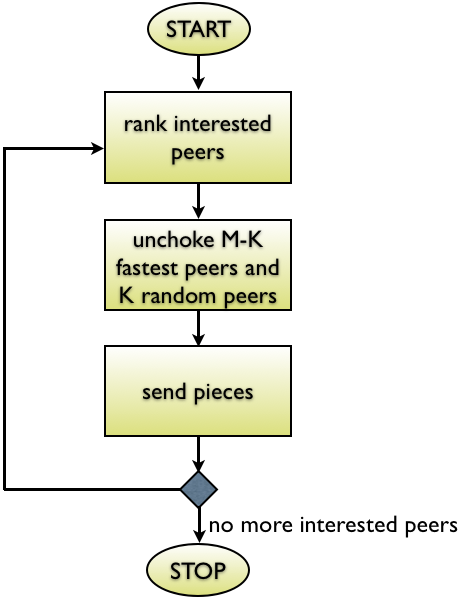}
\caption{The \textbf{Choke/Unchoke} strategy.}
\label{fig:chokeUnchoke}
\end{figure}

This solution (also sketched in Figure \ref{fig:chokeUnchoke}) follows a tit-for-tat (TFT) approach, since peer A is benevolent towards peer B, if the latter does the same, and vice versa.

\subsubsection{Examples} 
In BitTorrent~\cite{BitTorrent}, the \textbf{Choke/Unchoke} pattern is implemented with $M=5$, $K=1$, $T_1=10$s, $T_2=30$s. BitTorrent's enormous success suggests that TFT succeeds at inducing rational peers to share their resources. Moreover, TFT can be enforced without a centralized trusted infrastructure, thanks to its bilateral nature of TFT~\cite{Piatek07}.

BitTyrant~\cite{BitTyrant} is a BitTorrent client modified to reward strategic peers~\cite{Piatek2007}. BitTyrant's main principle is to carefully select contribution rates and peers, in order to maximize download per unit of upload bandwidth. The strategy of BitTyrant is based on policy (not protocol) client modifications, thus leaving unchanged the underlying BitTorrent protocol.

Tribler~\cite{Tribler} exploits social links between users behind peers to speed content discovery, recommendation and downloading up~\cite{Pouwelse07}. Tribler is based on the 2Fast protocol for collaborative downloading~\cite{Garba06}. In 2Fast, peers are either collectors or helpers, in a collaborative download process. A collector is a peer that wants to obtain a complete copy of a specific file. A helper, instead, is a peer that has been recruited by a collector, for downloading assistance. The collector and the helpers use the classical BitTorrent protocol to collaboratively download the file. A file chunk is selected for downloading based on RFA~\cite{BitTorrent}. However, before requesting and downloading a chunk, a helper asks the collector for its approval. If no other helper is already downloading the same chunk, the collector approves the helper's choice. 
The collector may optimize its download performance by dynamically choosing the best source from the set of helpers and other available peers in the BitTorrent network --- with preference for helpers and, in general, for peers with higher upload rates. 

\subsubsection{Consequences}
\textbf{Choke/Unchoke} supports data upload and download mutuality among peers.

\subsubsection{Related Patterns}
\textbf{Multisource Data Transfer}.

\subsection{\textbf{Reputation}}

\subsubsection{Context}
Not all peers put the same effort in the overlay network. Some peers are intrinsically resource-constrained, but there are also peers that wilfully limit their contribution (\textit{e.g.}, to save bandwidth). 

\subsubsection{Problem} 
How to reward the peers that contribute to the functioning of the network (\textit{e.g.}, by providing an adequate amount of upload bandwidth), while penalizing free riders? 

\subsubsection{Forces}
\begin{enumerate}
\item Rewarding good peers and penalizing bad peers is important for the survival of a P2P network.
\item Rewards and inflictions cannot be decided by a centralized entity, which would be a possible bottleneck and single point of failure for the system. 
\end{enumerate}

\subsubsection{Solution} 
\begin{figure}[h!]
\centering
\includegraphics[width=8cm]{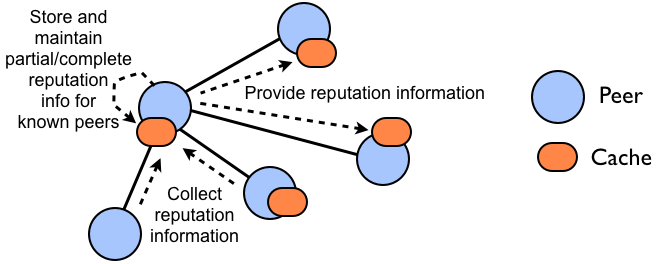}
\caption{The \textbf{Reputation} pattern.}
\label{fig:rep}
\end{figure}

Adopt a \textbf{Reputation} scheme (Figure \ref{fig:rep}). The \textbf{Reputation} of a peer is a numerical value representing its trustworthiness, based on past transaction activities. A \textbf{Reputation} scheme requires to store reputation information and to guarantee its integrity. 

\subsubsection{Variations}
\begin{figure}[h!]
\centering
\includegraphics[width=8cm]{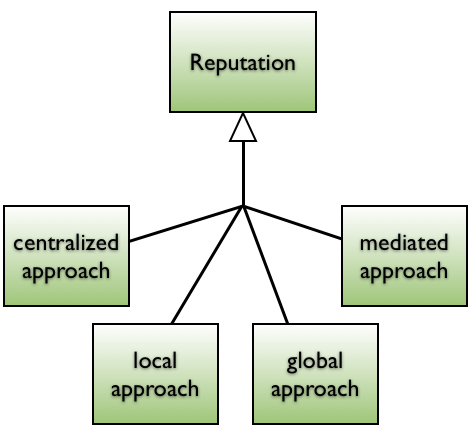}
\caption{Variations of the \textbf{Reputation} pattern.}
\label{fig:rep-var}
\end{figure}

Several \textbf{Reputation} approaches (Figure \ref{fig:rep-var}) exist:
\begin{enumerate} 
\item a stable and recognized peer stores and manage reputation information of all peergroup members (centralized approach);
\item every peer stores its experience against other peers, and when it is asked for reputation information about a particular peer, it provides a subjective answer (local approach, based on distributed credit);
\item reputation information is partitioned into several small pieces, which are stored throughout in the peergroup; that is, every peer equally manages a fraction of the whole reputation information (global approach);
\item only stable, recognized and highly-reputed peers are reputation maintainers (mediated approach).
\end{enumerate}

In the local solution, a distributed credit system is implemented by all peers. Transfer credit is locally maintained by all peers, to be considered when prospecting downloader will make a request for a data item. While obtaining a requested file, the receiving peer updates the credit associated to the sender, according to the amount of transferred data.

\subsubsection{Examples}
 In eMule~\cite{eMule1}, credit is calculated as the minimum of:
\begin{itemize}
\item $\frac{uploaded total \cdot 2}{downloaded total}$\\
When downloaded total is $0$, the expression evaluates to $10$.
\item $\sqrt{uploaded total + 2}$\\
When uploaded total is less then $1$ MB, the expression evaluates to $1$.
\end{itemize}
Upload/download amounts are measured in megabytes. In any case, the credit cannot exceed $10$ or be less than $1$.

Piatek \textit{et al.}~\cite{Piatek2008} proposed a one hop reputation protocol for P2P networks, which improves performance and incentives relative to BitTorrent. One hop reputations limit propagation to at most one level of indirection. Surprisingly, in most cases this limited propagation is sufficient to provide wide coverage.

\subsubsection{Consequences}
\textbf{Reputation} allows to reward the peers which contribute to the functioning of the network and to penalize free riders.

\section{Using the P2P-PL to implement overlay networks}
\label{implementation}

Working with P2P-PL requires the following procedure, where steps $1$ to $4$ are mandatory, while others depend on the purpose of the P2P system being designed.
\begin{enumerate}
\item Use the \textbf{Overlay Scheme} pattern to select the most appropriate P2P model that best matches your needs, depending on how information about shared resources must be placed.
\item If the selected P2P model is HM (hybrid) or DUM (decentralized unstructured), use the \textbf{Topology} and \textbf{Message Routing} patterns independently. Otherwise, if the P2P model is DSM (decentralized structured), use such patterns together --- in this case they are strictly correlated.
\item Use the \textbf{Bootstrapping} pattern to define the strategy that peers will adopt for joining the overlay network, by contacting already connected members.
\item Use the \textbf{Distributed State} pattern to implement a robust mechanism for identifying peers and resources within the network.
\item If the P2P system has to maintain consistency of resource descriptors among peers, in a decentralized fashion, use the \textbf{Information Consistency} pattern. 
\item If the P2P system has to guarantee the integrity, authenticity and availability of stored data, use the \textbf{Data Protection} pattern.
\item To prevent attacks based on open or loosely controlled membership, use the \textbf{Group Membership} pattern.
\item If the P2P system has to support content sharing or multimedia streaming, use the \textbf{Multisource Data Transfer} pattern to design a protocol for high-throughput data download.
\item If the P2P system has to support content sharing or multimedia streaming with a reasonable degree of upload and download mutuality, while penalizing free riders, use the \textbf{Choke/Unchoke} pattern.
\item To separate righteous peers from malicious peers, use the \textbf{Reputation} pattern.
\item Use distributed computing patterns (\textit{e.g.}, Remoting Patterns~\cite{Voelter04}, POSA2 patterns~\cite{POSA2}) to solve low-level issues.
\end{enumerate} 

In Table \ref{tab:useOfPatterns}, the presence of P2P-PL patterns in existing P2P architectures is reported. For example, let us consider BitTorrent (the traditional overlay network, not the DHT-based one). Nowadays, BitTorrent is one of the most relevant overlay networks in the file sharing domain. Its success is due to its robustness and efficiency, which derive from a careful design process. Not by chance, BitTorrent is probably the most P2P-PL -compliant P2P system, as described below.

\begin{itemize}
\item \textbf{Overlay Scheme}, in the HM variation. Indeed, BitTorrent relies on Web servers for storing .torrent static metainfo files, and on special agents (the \textit{trackers}) to create groups of peers interested in a same file.
\item \textbf{Bootstrapping}. In BitTorrent, joining an existing torrent requires the download of the related .torrent file from a torrent Web server. The .torrent file is provided to the peer along with the IP address of the tracker that manages the torrent. Then, the peer asks the tracker for a list of peers that are downloading and/or providing data chunks of the file associated to the torrent.
\item \textbf{Distributed State}. In BitTorrent, files and peers have unique identifiers. When started, every BitTorrent peer generates a $20$-byte identifier denoted as \textsf{peer\_id}. A new \textsf{peer\_id} is generated every time the peer is restarted. Every torrent descriptor contains a cryptographic hash for each piece of file. In this way, it is ensured that any alteration of the piece can be reliably detected. Having an authentic copy of the torrent descriptor, any peer can check the validity of the file it receives.
\item \textbf{Group Membership}. In BitTorrent, a peergroup that shares pieces of the same file is denoted as \textit{torrent swarm} (or just \textit{torrent}). Peers that own the complete file are denoted as \textit{seeds}, while peers that have a partial copy of the file and are trying to download the missing pieces are denoted as \textit{leechers}. Every peer maintains a list of known peers, denoted as \textit{peer set}. Thus, a torrent is like an ensemble of interconnected peer sets.
\item \textbf{Topology} In BitTorrent, the existence of a a link between two nodes depends on many factors. The number of upload connections may be limited not only by the protocol, but also by the user. To represent the topology of a group of downloaders, we use an undirected graph. For example, in figure~\ref{fig:groups-BT} three cases are illustrated:
\begin{enumerate}
\item $1$ seed for each file (torrent 1);
\item $1$ seed for $n$ files, thus $n$ possible groups (torrents 1 and 4, torrents 2 and 3);
\item $m$ seeds for the same file (torrent 4).
\end{enumerate}
Fauzie \textit{et al.} \cite{Fauzie2011} have shown that the topology of torrent swarms fits a power-law with exponential cutoff.
\begin{figure}[h!]
\centering
\includegraphics[width=10cm]{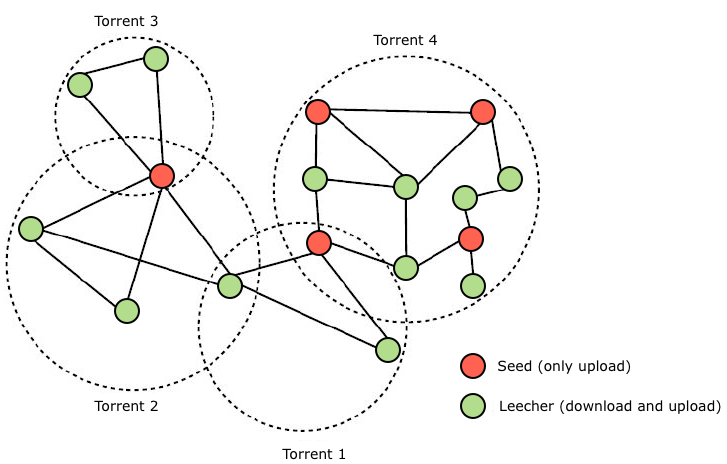}
\caption{Different kinds of torrents, in a BitTorrent network.}
\label{fig:groups-BT}
\end{figure}
\item \textbf{Message Routing}. In BitTorrent, only directly connected peers may exchange messages. Packets are not forwarded to destinations that are not listed in the peer set of the sender. 
\item \textbf{Multisource Data Transfer}. In BitTorrent, a peer can only send file pieces to the \textit{active peer set}, which is a subset of the peer set. File pieces are $256$ KB long and each piece is breaked into blocks of $16$ KB. For each peer in its peer set, every peer knows how pieces are distributed, thanks to frequent updates. Pieces are downloaded according to RFA \cite{Legout06}.
\item \textbf{Information Consistency}. In BitTorrent, .torrent files cannot be altered. If a new version of a file has to be shared, then a new .torrent must be generated and published on Web servers. Obsolete .torrent files, having no seeds for a long time, may be withdrawn by Web servers.
\item \textbf{Data Protection}. As previously observed, with reference to the \textbf{Distributed State} pattern, hashing is used to produce digests to quickly compare file pieces that claim to be copies of the same original version, provided by different peers. Moreover, both the header and the payload are encrypted, for each transferred piece. Only $60$-$80$ bits are used for the cipher, whose purpose is to obfuscate the stream as much as needed for making its detection very hard and resource-consuming \cite{Li2007}. 
\item \textbf{Choke/Unchoke}. It is BitTorrent's peer selection strategy, guaranteeing a reasonable degree of upload and download mutuality.
\end{itemize}
Only the \textbf{Reputation} P2P-PL pattern is not used in BitTorrent, for efficiency reasons. After all, \textbf{Multisource Data Transfer} and \textbf{Choke/Unchoke} are implemented in such a way that free riders are automatically penalized and at least one exemplar per file piece should be always available for download.

\begin{center}
\begin{table}
    \caption {Use of P2P-PL patterns in P2P architectures} 
    \label{tab:useOfPatterns} 
    \begin{tabular}{ | p{3cm} | p{0.8cm} | p{6cm} | }
    \hline
   Pattern & \# & P2P Systems \\ \hline \hline
   \textbf{Overlay Scheme} & n.a. & \textit{Every P2P system uses an \textbf{Overlay Scheme} variation} \\ \hline
   \textbf{Multisource Data Transfer} & 8 & BitTorrent, Gnutella, ChunkCast, CoolStreaming, PPLive, PPStream, UUSee, LayerP2P, ... \\ \hline
   \textbf{Message Routing} & 8 & Gnutella, MUTE, Freenet, JXTA, Chord, Pastry, Kademlia, BitTorrent, ... \\ \hline
   \textbf{Information Consistency} & 6 & JXTA, Chord, BitTorrent, eMule, Kademlia, Pastry, ... \\ \hline
   \textbf{Data Protection} & 5 & Freenet, OceanStore, eMule, BitTorrent, Skype, ... \\ \hline
   \textbf{Bootstrapping} & 3 & Skype, BitTorrent, eMule, ... \\ \hline
   \textbf{Topology} & 4 & Chord, Gnutella, Freenet, BitTorrent, ... \\ \hline
   \textbf{Choke/Unchoke} & 3 & BitTorrent, BitTyrant, Tribler, ...  \\ \hline
   \textbf{Distributed State} & 3 & JXTA, Chord, BitTorrent, ... \\ \hline
   \textbf{Group Membership} & 2 & BitTorrent, JXTA, ... \\ \hline
   \textbf{Reputation} & 1 & eMule, ... \\ \hline
   \end{tabular}
\end{table}
\end{center}

\section{Concluding Remarks}
\label{conclusions}

In this paper, we presented P2P-PL, a pattern language for robust P2P overlay networks. We put a lot of effort in focusing on P2P-specific problems and solutions, separating core P2P patterns (which are mandatory for any P2P system) from more specific P2P patterns. In doing this, we took into account the rich existing literature about patterns for distributed computing, that need to be used after the P2P-PL (according to the top-down approach we suggest). The resulting pattern language should be helpful for P2P system designers and developers, who will find a well-organized collection of best practices and examples, together with a stepwise design procedure.

\section{Glossary}

In Table \ref{tab:glossary}, a glossary is provided, gathering terms that are adopted in multiple patterns within P2P-PL.

\begin{center}
\begin{table}
    \caption {Glossary} 
    \label{tab:glossary} 
    \begin{tabular}{ | p{3cm} | p{9cm} | }
    \hline
    Term & Meaning \\ \hline \hline
    Overlay network & Built on top of an existing computer network, an overlay network is a virtual network of nodes and logical links, with the purpose to support specific services. Peer-to-peer systems are overlay networks running on top of the Internet, which is an overlay network that runs on top of several physical networks. \\ \hline
    Network topology & The arrangement of the nodes and links of a network is denoted as network topology. The physical topology of a network refers to devices and cables or radio links, while the logical topology shows how data packets flow within the network, independently of its physical design. \\ \hline
    Node & A physical or virtual machine connected to other components of the same type, within a network. \\ \hline
    Resource & Any entity which can be advertised in a network (\textit{e.g.} nodes, files, services, CPU cores, storage spaces, etc.). \\ \hline
    Correctness & The system does exactly what it is intended to do, no more, no less. \\ \hline
    Efficiency & The system operates avoiding waste of resources (memory, CPU cycles, bandwidth, data transmissions, etc.). \\ \hline
    Robustness & The system continues to operate despite abnormal input, operations, etc. \\ \hline
    \end{tabular}
\end{table}
\end{center}






\end{document}